\documentclass[preprintnumbers,showkeys]{revtex4}
\usepackage{amsmath,amssymb,graphics,epsfig,subfigure}
\usepackage{color}
\usepackage{float}
\usepackage{multirow}
\usepackage{booktabs}

\newcommand{\club}{\clubsuit}
\newcommand{\spad}{\spadesuit}
\newcommand{\loze}{\blacklozenge}
\usepackage{hyperref}
\hypersetup{colorlinks=true,linkcolor=blue,citecolor=magenta}
\begin{document}
\renewcommand{\baselinestretch}{1.3}

\title{Novel topological phenomena of timelike circular orbits for charged test particles}

\author{Xu Ye, Shao-Wen Wei\footnote{Corresponding author. E-mail: weishw@lzu.edu.cn}}

\affiliation{$^{1}$Key Laboratory of Quantum Theory and Applications of MoE, Lanzhou University, Lanzhou 730000, China\\
$^{2}$Lanzhou Center for Theoretical Physics, Key Laboratory of Theoretical Physics of Gansu Province, School of Physical Science and Technology, Lanzhou University, Lanzhou 730000, People's Republic of China,\\
$^{3}$Institute of Theoretical Physics $\&$ Research Center of Gravitation,
Lanzhou University, Lanzhou 730000, People's Republic of China}

\begin{abstract}
The topological approach has recently been successfully employed to investigate timelike circular orbits for massive neutral test particles. The observed vanishing topological number implies that these timelike circular orbits occur in pairs. However, the behavior of charged test particles in this regard remains unexplored. To address this issue, our study focuses on examining the influence of particle charge on the topology of timelike circular orbits within a spherically symmetrical black hole spacetime holding a nonvanishing radial electric field. We consider four distinct cases based on the charges of the particle and the black hole: unlike strong charge, unlike weak charge, like weak charge, and like strong charge. For each case, we calculate the corresponding topological number. Our results reveal that when the charge is large enough, the topological number takes a value of -1 instead of 0, which differs from the neutral particle scenario. Consequently, in cases of small charges, the timelike circular orbits appear in pairs, whereas in cases of larger charges, an additional unstable timelike circular orbit emerges. These findings shed light on the influence of the particle charge on the topological properties and number of timelike circular orbits.
\end{abstract}



\maketitle

\section{Introduction}

Topology has emerged as a promising approach in the exploration of the underlying information of black hole physics. By disregarding the local properties of the black hole system, topology offers insights into its global properties. Notably, this approach has successfully been employed to study the light ring, timelike circular orbits, and thermodynamics of black holes, leading to the discovery of even more intriguing properties \cite{Cunha2017,Cunha2020azh,Wei2022mzv,Wei2022dzw}.

In the recent study \cite{Cunha2017}, Cunha, Berti, and Herdeiro constructed a characteristic vector field that relates to the topology of light ring (LR). Through this construction, they discovered that the zero points of the vector field correspond exactly to the LRs on the equatorial plane in stationary axisymmetric spacetimes. Moreover, they established a universal relationship between the stability of the LRs and the winding number of its zero points for these ultracompact objects. The LRs always come in pairs with one of them being stable. Subsequently, this topological method was extended to the rotating black holes \cite{Cunha2020azh}. In contrast to ultracompact objects, stationary, axisymmetric, asymptotically flat black holes must have at least one unstable LR outside the event horizon.

In addition to the LRs, the circular orbit of the massive particles is another crucial aspect to consider. In Ref. \cite{Delgado2021jxd}, the authors highlighted the significance of the timelike circular orbits (TCOs). The results show that there exists a close relation between the LRs and TCOs. For example, the energy and angular momentum of the TCOs diverge if they coincide with the LRs. Furthermore, the radial stability of TCOs can be effectively tested near the LRs. Specifically, in the vicinity of a stable LR, TCOs with a slightly larger radius exhibit radial instability. Conversely, in the vicinity of an unstable LR, TCOs with a slightly smaller radius are stable. Importantly, no TCOs can survive in the radial region between a stable LR and an unstable LR.

Given the relationship between LRs and TCOs, one might wonder whether the topological approach can be extended to TCOs. However, unlike LRs, the location of TCOs is closely dependent on the energy and angular momentum of the massive particles within a given spacetime. This particular characteristic makes the LR approach unsuitable for analyzing TCOs. Fortunately, we have discovered a solution to this issue in Ref. \cite{Wei2022mzv}. It is revealed that if the angular momentum of the TCO is known, a vector can be defined in such a way that its zero points precisely correspond to the locations of TCOs. This discovery allows for the establishment of a well-defined and effective topology to characterize TCOs. Our study reveals that the total topological number of the TCOs for a generic stationary axisymmetric black hole is zero. This observation implies that the stable and unstable TCOs either do not exist or occur in pairs since they possess opposite topological numbers. Significantly, when the angular momentum of the particles is selected as a control parameter, the ISCO acts as a bifurcation point \cite{Fu2000pb}. Even in black hole solutions with the scalar hair and quasitopological terms, more diverse topological configurations for TCOs are exhibited. Nevertheless, the total topological number remains unchanged \cite{Ye2023gmk}.

Topology also finds important applications in black hole thermodynamics, particularly in the exploration of critical points and phase transitions. Recent studies have revealed the existence of two distinct critical points, indicating different phase structures of black holes. To differentiate between these critical points, a topology is constructed in Ref.  \cite{Wei2021vdx} by using Duan's topological current $\phi$-mapping theory \cite{Duan1984ws}. By calculating the winding number, the critical points can be classified as either novel or conventional. Additionally, black hole systems can be categorized into different topological classes based on the topological number. Furthermore, these study has been extended to other black hole backgrounds, leading to the discovery of even more intriguing topological properties \cite{Sadeghi2023dsg,Hazarika2023iwp,Hazarika2024cpg}. Soon afterwards, a fascinating new idea was proposed. By utilizing the generalized off-shell free energy \cite{Wei2022dzw}, black hole solutions can be treated as topological defects in the thermodynamic parameter space. Each black hole is then associated with a topological charge known as the winding number. Positive or negative winding number indicates that the black hole is thermodynamical stable or unstable. Moreover, by summing these winding numbers at a given temperature, one can obtain a global topological number. From a thermodynamic perspective, black hole solutions can be classified into different classes. For instance, the four-dimensional Schwarzschild, charged, and charged AdS black holes have total topological numbers of -1, 0, and 1, respectively. This suggests that these black holes belong to different topological classes. Subsequently, this topological approach has been extended to explore massive gravity, Lovelock gravity, and other modified gravities \cite{Chen2023ddv, Bai2022klw, Hung2023ggz}.

Recently, the dynamics of massive charged particles surrounding compact objects has gained significant attention, particularly in the field of relativistic astrophysics. The motion of these charged particles is directly influenced by the gravitational properties of the sources, allowing us to extract valuable information about the compact objects through the study of the geometric and physical characteristics of their motion trajectories. In Ref. \cite{Pugliese2011py}, Pugliese, Quevedo, and Ruffini examined the circular motion of charged test particles in Reissner-Nordstr\"om (RN) black holes. Their results demonstrated that by analyzing the geometric structure of stable accretion disks composed solely of charged particles moving along the circular orbits, it is possible to clearly distinguish the black holes and naked singularities. Furthermore, they provided a comprehensive classification of circular orbits by varying the charge-to-mass ratios \cite{Pugliese2017xfa}. Notably, other studies have also explored the behavior of charged particles in the black hole background with magnetic fields \cite{Wald1974np, Grunau2010gd, Shaymatov2021qvt, Garcia2013zud}.

For the circular orbits, previous topological studies have mainly focused on the photons and massive neutral particles, while leaving the case of charged particles unexplored. However, if black holes possess electric charge, the presence of Coulomb interaction presents the potential case for a balance between gravity and electromagnetism. This leads to a distinct structure of TCO. Another notable example is the violation of the chaos bound by the TCOs of charged particles in RN black holes \cite{Zhao2018wkl,Lei2021koj}. Consequently, one can expect the emergence of novel topological configurations. Whether topological phase transitions occur is also a valuable question to explore. In this paper, we focus on these issues and demonstrate that when the Coulomb interaction dominates over gravity, the topological number of the TCOs changes from 0 to -1, indicating an underlying topological phase transition.

This paper is organized as follows. In Sec. \ref{sec2}, we solve the radial effective potential for charged particles in a generic spherically symmetric spacetime with a radial electric field, and then examine the asymptotic behavior of the vector field at the boundaries of the spacetime. The total topological number is obtained. In Sec. \ref{sec3}, we focus specifically on the RN black holes. Sec. \ref{sec4} is devoted to the study of the topology of the TCOs, where four cases characterized by distinct charge-to-mass ratios are considered. Finally, we summarize and discuss our results in Sec. \ref{sec5}. Throughout this paper, we adopt the geometrized unit system where $c=\hbar=G=1$.

\section{Generic spherically symmetrical charged black hole}\label{sec2}

In this section, we would like to explore the geodesics of charged particles and derive the effective potential for the radial motion, then construct the corresponding vector of the timelike circular orbits. By examining the behavior of the vector field at the boundaries, the topological number shall be obtained for a spherically symmetrical spacetime.

\subsection{Effective potential and timelike circular orbit}

The Lagrangian of a charged test particle under the curved spacetime is \cite{Chandrasekhar1985kt}
\begin{equation}\label{Lag}
    \mathcal{L}=\frac{1}{2}g_{\alpha \beta}\dot{x}^\alpha \dot{x}^\beta+q A_\alpha \dot{x}^\alpha.
\end{equation}
The parameter $q$ represents the charge per unit mass of the particle, which can have either a positive or negative value. The dot denotes the derivative with respect to the affine parameter.

Assuming that the electromagnetic potential has only the $t$ component and takes the following form
\begin{equation}\label{Afp}
    A_\mu=\left(\frac{Q}{r},\, 0,\, 0,\, 0\right),
\end{equation}
under which, one can derive the radial electric field $E_r=- \nabla A_t=Q/r^2$, where $Q$ represents the electric charge of the black hole solution.

In general, the line element of a spherically symmetrical black hole spacetime reads
\begin{equation}\label{LE}
    ds^2=-f(r) dt^2+\frac{1}{f(r)} dr^2+ r^2 d \theta^2+r^2 \sin^2 \theta d\phi^2,
\end{equation}
where metric function $f(r)$ only depends on the radial coordinate $r$.

By employing Eqs. (\ref{Afp}) and (\ref{LE}), we can reformulate the Lagrangian (\ref{Lag}) as
\begin{equation}
    \mathcal{L}= \frac{1}{2}\left(-f(r) \dot{t}^2+\frac{1}{f(r)} \dot{r}^2+ r^2 \dot{\theta}^2+r^2 \sin^2 \theta \dot{\phi^2}\right)+ \frac{q Q}{r} \dot{t}.
\end{equation}

Due to the spherical symmetry of spacetime, two Killing vectors, namely $\partial/\partial_t$ and $\partial/\partial_\phi$, correspond to the conservation laws of energy $E$ and angular momentum $L$ for the charged particles, serving as the conjugate momenta of the coordinates $t$ and $\phi$, respectively
\begin{align}
    &P_t := \frac{\partial \mathcal{L}}{\partial \dot{t}}= \frac{qQ}{r}-f(r) \dot{t}=E, \label{EE}\\
    &P_\phi := \frac{\partial \mathcal{L}}{\partial \dot{\phi}}= r^2\sin^2\theta \dot{\phi }=L \label{LL}.
\end{align}
Therefore, we can easily obtain $\dot{t}$ and $\dot{\phi}$
\begin{equation}\label{Tdot}
    \dot{t}=\frac{qQ-Er}{f(r)r}, \qquad \dot{\phi}=\frac{L\csc^2 \theta}{r^2}.
\end{equation}
For the timelike geodesics, we have $g_{\mu\nu}\dot{x}^{\mu}\dot{x}^{\nu}=-1$ leading to
\begin{equation}\label{norm1}
    f(r) \dot{t}^2-\frac{1}{f(r)} \dot{r}^2-r^2 \dot{\theta}^2-r^2 \sin^2 \theta^2 \dot{\phi}^2=1.
\end{equation}
Substituting Eq. (\ref{Tdot}) into (\ref{norm1}), it is easy to obtain
\begin{equation}
    \frac{\dot{r}^2}{f(r)}+r^2 \dot{\theta}^2+V_{eff}=0,
\end{equation}
where the effective potential reads
\begin{equation}\label{Veff}
    V_{eff}:=1-\frac{(qQ-E r)^2}{f(r) r^2}+\frac{L^2\csc^2\theta}{r^2}.
\end{equation}
In particular, it can be further factored into
\begin{equation} \label{Veff2}
    V_{eff}= \frac{-1}{f(r)}(E-E_1)(E-E_2),
\end{equation}
where
\begin{align}
    &E_1= \frac{qQ}{r}+\frac{\sqrt{f(r)\csc ^2(\theta ) \left(2 L^2-r^2 \cos (2 \theta )+r^2\right)}}{\sqrt{2}r}, \label{eqE1}\\
    &E_2= \frac{qQ}{r}-\frac{\sqrt{f(r)\csc ^2(\theta ) \left(2 L^2-r^2 \cos (2 \theta )+r^2\right)}}{\sqrt{2}r}. \label{eqE2}
\end{align}
It should be noted that the conditions for a TCO determined by $V_{\text{eff}}(r) = 0$ and $V_{\text{eff}}'(r) = 0$, can be alternatively expressed in terms of $E_1$ and $E_2$
\begin{equation} \label{eqcond}
    E=E_1 \; \wedge \; E_1'=0, \quad\text{or}\quad  E=E_2 \; \wedge \; E_2'=0.
\end{equation}
The prime denotes the derivative with respect to the radial coordinate $r$.

\subsection{Energy and angular momentum}

Solving $E_1'=0$, we can obtain the angular momentum
\begin{align}
    L_{1A}&=\pm\sqrt{\frac{2 f(r) \left(r^3 f'(r)+q^2 Q^2\right)-2 \sqrt{q^2 Q^2 f(r)^2 \left(-2 r^3 f'(r)+4 r^2 f(r)+q^2 Q^2\right)}- r^4 f'(r)^2}{\left(r f'(r)-2 f(r)\right)^2}}, \label{L1A}\\
    L_{1B}&=\pm\sqrt{\frac{2 f(r) \left(r^3 f'(r)+q^2 Q^2\right)+2 \sqrt{q^2 Q^2 f(r)^2 \left(-2 r^3 f'(r)+4 r^2 f(r)+q^2 Q^2\right)}- r^4 f'(r)^2}{\left(r f'(r)-2 f(r)\right)^2}} \label{L1B}.
\end{align}
The presence of the upper and lower signs $\pm$ in $L_{1A}$ and $L_{1B}$ is a consequence of the spherical symmetry of the background spacetime. This symmetry implies that the TCO remains unchanged under the reflection transformation $L\to-L$.

Putting \eqref{L1A} and \eqref{L1B} back into \eqref{eqE1}, the corresponding energy are
\begin{align}
    E_{1A}&=\frac{qQ}{r}+\frac{f(r)}{r}\sqrt{\frac{4 r^2 f(r)-2 \sqrt{q^2 Q^2 \left(-2 r^3 f'(r)+4 r^2 f(r)+q^2 Q^2\right)}-2r^3 f'(r) +2q^2 Q^2}{\left(r f'(r)-2 f(r)\right)^2}}, \label{E1A}\\
    E_{1B}&=\frac{qQ}{r}+ \frac{f(r)}{r} \sqrt{\frac{4 r^2 f(r)+2\sqrt{q^2 Q^2 \left(-2 r^3 f'(r)+4 r^2 f(r)+q^2 Q^2\right)} - 2r^3 f'(r) +2q^2 Q^2}{\left(r f'(r)-2 f(r)\right)^2}} . \label{E1B}
\end{align}
Adopting the similar calculation, we obtain the angular momentum from $E_2'=0$. However, the results are identical with Eqs. \eqref{L1A} and \eqref{L1B}
\begin{equation}
    L_{2A}=L_{1A}, \qquad L_{2B}=L_{1B}.
\end{equation}
Substituting $L_{2A}$ and $L_{2B}$ back into \eqref{eqE2}, the distinct TCO energy takes the form
\begin{align}
    E_{2A}&=\frac{qQ}{r}- \frac{f(r)}{r} \sqrt{\frac{4 r^2 f(r)-2 \sqrt{q^2 Q^2 \left(-2 r^3 f'(r)+4 r^2 f(r)+q^2 Q^2\right)}-2r^3 f'(r)+2q^2 Q^2}{\left(r f'(r)-2 f(r)\right)^2}}, \\
    E_{2B}&=\frac{qQ}{r}- \frac{f(r)}{r}\sqrt{\frac{4 r^2 f(r)+2\sqrt{q^2 Q^2 \left(-2 r^3 f'(r)+4 r^2 f(r)+q^2 Q^2\right)} -2r^3f'(r) +2q^2 Q^2}{\left(r f'(r)-2 f(r)\right)^2}}.
\end{align}
As a result, within this static spherically symmetric spacetime characterized by a radial electric field, there exist at most four distinct combinations of energy and angular momentum for the massive charged test particles: ($E_{1A}$, $L_{1A}$), ($E_{1B}$, $L_{1B}$), ($E_{2A}$, $L_{2A}$), and ($E_{2B}$, $L_{2B}$). This situation is more complicated compared to the scenario for the neutral particles, where only a single solution is present, as shown in Ref. \cite{Ye2023gmk}. However, it is important to note that the four solution sets are not uniformly well behaved for a specific black hole background. In certain regions of the parameter space, they are valid solutions, while in other regions, they are not. This gives significant challenges to our investigation. To address this issue, it becomes necessary to reformulate the solution. Fortunately, the solutions corresponding to the four sets can be expressed in the following form:
\begin{align}
    &E_{t1}= -\frac{\sqrt{f(r)^2 \left(-2 r^3 f'(r)+4 r^2 f(r)+q^2 Q^2\right)}+q Q\left( f(r)- r f'(r)\right)}{r^2 f'(r)-2r f(r)},\label{Et1}\\
    &L_{t1}= \sqrt{\frac{2 f(r) \left(r^3 f'(r)+q^2 Q^2\right)-r^4f'(r)^2-2 q Q f(r)\sqrt{ -2 r^3 f'(r)+4 r^2 f(r)+q^2 Q^2}}{ \left( 2 f(r)-r f'(r) \right)^2}};
\end{align}
\begin{align}
    &E_{t2}= \frac{\sqrt{f(r)^2 \left(-2 r^3 f'(r)+4 r^2 f(r)+q^2 Q^2\right)}+q Q \left( r f'(r)- f(r) \right) }{r^2 f'(r)-2r f(r)},\\
    &L_{t2}=\sqrt{\frac{2 f(r) \left(r^3 f'(r)+q^2 Q^2\right)-r^4f'(r)^2+2 q Q f(r)\sqrt{-2 r^3 f'(r)+4 r^2 f(r)+q^2 Q^2}}{ \left(2 f(r)-r f'(r)\right)^2 }}; \label{Lt2}
\end{align}
\begin{align}
    &E_{t3}= -\frac{\sqrt{f(r)^2 \left(-2 r^3 f'(r)+4 r^2 f(r)+q^2 Q^2\right)}+q Q\left( f(r)-r f'(r)\right)}{ r^2 f'(r)-2 r f(r)},\\
    &L_{t3}= -\sqrt{\frac{2 f(r) \left(r^3 f'(r)+q^2 Q^2\right)-r^4f'(r)^2-2 q Q f(r) \sqrt{-2 r^3 f'(r)+4 r^2 f(r)+q^2 Q^2}}{ \left( 2 f(r)-r f'(r)\right)^2 }};
\end{align}
\begin{align}
    &E_{t4}= \frac{\sqrt{f(r)^2 \left(-2 r^3 f'(r)+4 r^2 f(r)+q^2 Q^2\right)}+q Q\left( r f'(r)- f(r)\right) }{ r^2 f'(r)-2r f(r)},\\
    &L_{t4}= -\sqrt{\frac{2 f(r) \left(r^3 f'(r)+q^2 Q^2\right)-r^4f'(r)^2+2 q Q f(r)\sqrt{ -2 r^3 f'(r)+4 r^2 f(r)+q^2 Q^2}}{\left( 2 f(r)-r f'(r)\right)^2 }} \label{Lt4}.
\end{align}
Obviously, we find the following relation
\begin{eqnarray}
    &E_{t1}=E_{t3}\quad \text{and} \quad  L_{t1}=-L_{t3},\\
    &E_{t2}=E_{t4}\quad \text{and} \quad L_{t2}=-L_{t4}.
\end{eqnarray}
Considering the spherical symmetry of the spacetime background, it suffices to analyze the solutions for the TCOs with positive angular momentum. Consequently, we obtain two distinct sets of solutions: $E_{t1}$ and $L_{t1}$, as well as $E_{t2}$ and $L_{t2}$.

Subsequently, we turn our attention to the ISCO as it is closely related to the stability of the TCOs. In general, if the radius of the TCO is greater than $r_{\text{ISCO}}$, the TCO is radially stable, otherwise, it is unstable. For a generalization, we would like to refer to Ref. \cite{Delgado2021jxd} if MSCO appears. The ISCO is usually determined by
\begin{equation}\label{ad}
    V_{eff}(r)\big|_{r=r_{ISCO}}=0\;\;\wedge\;\; V_{eff}'(r)\big|_{r=r_{ISCO}}=0 \;\; \wedge\;\; V_{eff}''(r)\big|_{r=r_{ISCO}}=0.
\end{equation}
Solving the third condition (\ref{ad}), we have
\begin{equation}\label{IScons}
    \begin{split}
    V_{eff}''(r)=& \frac{-2 r f'(r)^2 (q Q-E r)^2+ f(r) (q Q-E r) \left(r f''(r) (q Q-E r)-4 q Q f'(r)\right)}{r^3 f(r)^3}\\
    &  +\frac{6 L^2 f(r)+2 q Q  (2 E r-3 q Q)}{r^4 f(r)} = 0.
    \end{split}
\end{equation}
By substituting $E_{t1}$ and $L_{t1}$, as well as $E_{t2}$ and $L_{t2}$, into (\ref{IScons}), we derive two equations associated with the ISCO
\begin{eqnarray} \label{IS1}
    &f(r)^2 \left(2r^4 f''(r)+6 r^3 f'(r)-2q^2 Q^2\right)-q Q f(r) \left(r^2 f''(r)-2r f'(r)+2 f(r)\right) \sqrt{-2 r^3 f'(r)+4 r^2 f(r)+q^2 Q^2} \nonumber\\
    &\quad +r^2 f'(r)^2 \left(2 r^3 f'(r)-q^2 Q^2\right)+r f(r) \left(q^2 Q^2 r f''(r)-7 r^3 f'(r)^2+f'(r) \left(2 q^2 Q^2-r^4 f''(r)\right)\right)=0,\\
    \label{IS2}
    &f(r)^2 \left(2 r^4 f''(r)+6 r^3 f'(r)-2 q^2 Q^2\right)+q Q f(r) \left( r^2 f''(r)-2r f'(r)+2 f(r)\right) \sqrt{-2 r^3 f'(r)+4 r^2 f(r)+q^2 Q^2}\nonumber\\
    &\quad +r^2 f'(r)^2 \left(2 r^3 f'(r)-q^2 Q^2\right)+r f(r) \left(q^2 Q^2 r f''(r)-7 r^3 f'(r)^2+f'(r) \left(2 q^2 Q^2-r^4 f''(r)\right)\right)=0.
\end{eqnarray}

The above equations indicate that the ISCO relies on the black hole charge $Q$, the charge-to-mass ratio $q$ of the timelike particle, and the metric function $f(r)$. If a specific $f(r)$ is given, we shall obtain the radius of the ISCO in terms of $Q$, $q$, and other black hole parameters. Henceforth, in the subsequent discussion, we will refer the solutions to Eqs. (\ref{IS1}) and (\ref{IS2}) as $r_{\text{ISCO}}(E_{t1}, L_{t1})$ and $r_{\text{ISCO}}(E_{t2}, L_{t2})$, respectively.

\subsection{Vector and topological number}

In this subsection, our aim is to construct two vectors $\vec{\phi}_1=(\phi^r_1,\phi^\theta_1)$ and $\vec{\phi}_2=(\phi^r_2,\phi^\theta_2)$ separately using $E_1$ and $E_2$, such that their zero points exactly correspond to the TCOs. Treating these zero points as topological defects, we can assign a local topological charge to each of them, known as the winding number. By summing up all these topological charges, we obtain a topological number that characterizes the TCOs. This topological number reveals the global characteristic of the spacetime under consideration.

At first, we examine the scenario where the vector field is constructed using $E_1$. Then, the radial and angular components of vector $\vec{\phi}_1$ are defined as \cite{Wei2022mzv}
\begin{equation}\label{Vec}
    \vec{\phi}_1=\left(\phi^r_1,\; \phi^\theta_1 \right) :=\left( \frac{\partial_r E_1}{\sqrt{g_{rr}}},\; \frac{\partial_\theta E_1}{\sqrt{g_{\theta \theta}}}\right),
\end{equation}
where $g_{rr}=\frac{1}{f(r)}$ and $g_{\theta \theta}=r^2$.

Employing with Eq. (\ref{eqE1}), the explicit expressions of $\phi^r_1$ and $\phi^\theta_1$ are given by
\begin{align}
    \phi^r_1=&\;\frac{\sqrt{2} \csc(\theta ) f'(r) \sqrt{2 L^2-r^2 \cos (2 \theta)+r^2}}{4 r }-\frac{ q Q  \sqrt{f(r) }}{ r^2 } - \frac{ \sqrt{2}L^2 f(r) \csc(\theta)}{ r^2 \sqrt{ 2 L^2-r^2 \cos (2 \theta )+r^2}}, \label{phir1} \\
    \phi^\theta_1=&\;-\frac{\sqrt{2f(r)} L^2 \cot (\theta ) \csc(\theta )}{r^2 \sqrt{ 2 L^2-r^2 \cos (2 \theta )+r^2}}. \label{phith1}
\end{align}
Furthermore, when combined with the conditions (\ref{eqcond}), it becomes evident that the zero points of the vector, i.e., $\phi^r_1 = \phi^\theta_1 = 0$, are located at $r = r_{\text{tco}}$ and $\theta = \pi/2$.

Next, we aim to investigate the boundary behaviors of the vector $\vec{\phi}_1=(\phi^r_1,\phi^\theta_1)$ and analyze the topological charge in the context of this generic spherically symmetric spacetime. The considered spacetime boundary is denoted as $\partial\Sigma$, which encompasses the region $\Sigma$
\begin{equation} \label{Reg}
    \Sigma=\left\{ I_r \times I_\theta \subset \mathbb{R}^2  \bigg| I_r=\left[r_h,\; \infty \right) \subset \mathbb{R}  , I_\theta = \left[0,\; \pi \right]  \subset \mathbb{R} \right\}.
\end{equation}
Therefore, the boundary $\partial \Sigma$ consists of four segments, the event horizon $r=r_h$, spatial infinity $r=\infty$, the polar axis at $\theta=0$ and $\pi$.

\textbf{i) Asymptotic limit ($r\to \infty$)}

We assume that spacetime is asymptotical Schwarzschild case, which gives
\begin{equation}
    f(r) =1-\frac{2M}{r}+\mathcal{O}\left(r^{-2}\right).
\end{equation}
as $r\to \infty$. Here $M$ measures the black hole mass and is always positive.

By performing Taylor expansion of $\vec{\phi}_1$ near $r=\infty $, we obtain the following results
\begin{align}
    &\phi^r_1=\frac{M-q Q}{r^2}+\mathcal{O}\left(r^{-3}\right),\\
    &\phi^\theta_1=-\frac{L^2 \cot (\theta ) \csc ^2(\theta)}{r^3}+\mathcal{O}\left(r^{-4}\right).
\end{align}
The argument of the vector reads
\begin{equation}
    \arctan\left(\frac{\phi^\theta_1}{\phi^r_1}\right)=\arctan\left(-\frac{L^2 \cot (\theta ) \csc ^2(\theta )}{r (M-q Q)}\right)\bigg|_{r\to\infty}=0.
\end{equation}
This observation suggests that $\phi^\theta_1$ is significantly smaller than $\phi^r_1$, indicating that the vector field $\vec{\phi}_1$ predominantly aligns horizontally along the $r$-direction. Additionally, we observe that the sign of $\phi^r_1$ depends on the difference between $M-qQ$. Consequently, two distinct scenarios arise concerning the direction of the vector field
\begin{equation}\label{infB}
\begin{cases}
    qQ<0\;\;\text{or}\;\; 0<qQ<M, & \vec{\phi}_1\;\; \text{points to the right};\\
    0<M<qQ, &  \vec{\phi}_1\;\; \text{points to the left}.
\end{cases}
\end{equation}

\textbf{ii) Horizon limit ($r\to r_h$)}

Near the event horizon $r=r_h$, the metric function $f(r)$ can be expanded as
\begin{equation}
    f(r)=f\left(r_h\right)+\left(r-r_h\right) f'\left(r_h\right)+\frac{1}{2}\left(r-r_h\right){}^2 f''\left(r_h\right)+\mathcal{O}\left(\left(r-r_h\right){}^3\right).
\end{equation}
Note that $f(r_h)=0$ and $f'(r_h)>0$. By making use of it, we have
\begin{align}
    &\phi^r_1=\frac{\csc(\theta ) f'(r_h) \sqrt{r_h^2+2 L^2-r_h^2 \cos (2 \theta )}}{2 \sqrt{2} r_h }+\mathcal{O}\left(\left(r-r_h\right)\right), \\
    &\phi^\theta_1=-\frac{L^2 \sqrt{2 f(r_h)} \cot (\theta ) \csc(\theta )}{r_h^2 \sqrt{r_h^2+2 L^2-r_h^2 \cos (2 \theta )}}+\mathcal{O}\left(\left(r-r_h\right)\right).
\end{align}

Since the angle between $\phi^\theta_1$ and $\phi^r_1$ is determined by the ratio $\phi^\theta_1/\phi^r_1$, we just focus on calculating this ratio as follows
\begin{equation}
    \frac{\phi^\theta_1}{\phi^r_1}=\frac{4 L^2 \cot (\theta ) \sqrt{f(r_h)} }{r_h f'\left(r_h\right) \left(r_h^2 \cos (2 \theta)-r_h^2-2 L^2\right) }\bigg|_{f(r_h)\to 0}=0.
\end{equation}
Hence, considering that $\phi^\theta_1$ is significantly smaller than $\phi^r_1$ and $\phi^r_1\propto \sqrt{r_h^2\left(1-\cos(2 \theta)\right)+2L^2}>0$, we can conclude that the vector field $\vec{\phi}_1$ on the horizon $r_h$ points to the right.

\textbf{iii) Axial limit ($\theta\to 0$ and $\pi$)}

Near $\theta\to 0$, $\phi^r_1$ and $\phi^\theta_1$ can be expanded as
\begin{align}
    &\phi^r_1=\frac{|L| \left(r f'(r)-2 f(r)\right)}{2 \theta  r^2}+\mathcal{O}\left(\theta ^0\right), \\
    &\phi^\theta_1=-\frac{|L|\sqrt{f(r)}}{\theta ^2 r^2}+\mathcal{O}\left(\theta ^0\right).
\end{align}
As a result, we have
\begin{equation}
    \frac{\phi^\theta_1}{\phi^r_1}=-\frac{2 \sqrt{f(r)}}{\theta \left(r f'(r)-2 f(r)\right)} \bigg|_{\theta\to 0}=\infty.
\end{equation}
Combining with $\phi^\theta_1\gg\phi^r_1$ and $\phi^\theta_1<0$, we conclude that the direction of vector field is downward at $\theta=0$.

Performing the similar calculation, we obtain
\begin{align}
    &\phi^r_1=-\frac{|L| \left(r f'(r)-2 f(r)\right)}{2 (\theta -\pi) r^2}+\mathcal{O}\left((\theta -\pi )^0\right),\\
    &\phi^\theta_1= \frac{|L|\sqrt{f(r)}}{(\theta -\pi )^2 r^2}+\mathcal{O}\left((\theta -\pi )^0\right).
\end{align}
near $\theta=\pi$. The corresponding ratio reads
\begin{equation}
    \frac{\phi^\theta_1}{\phi^r_1}=-\frac{2 \sqrt{ f(r)}}{(\theta -\pi ) \left(r f'(r)-2 f(r)\right)}\bigg|_{\theta\to\pi}=\infty.
\end{equation}
Further considering $\phi^\theta_1>0$, the direction of vector field is upward at $\theta=\pi$.

\begin{figure}[htb]
    \centering
    \includegraphics[width=7cm]{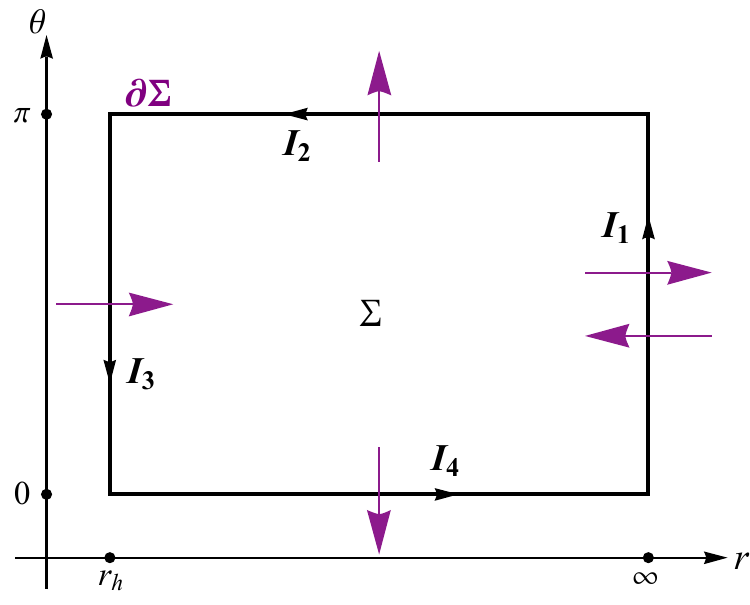}
    \caption{The boundary behaviors of vector field $\vec{\phi}_1=(\phi^r_1,\phi^\theta_1)$ constructed from $E_1$ on ($r,\,\theta$) plane. The purple arrows represent the direction of vector field, which are uniquely determined at $r\to r_h$ and $\theta\to 0$ or $\pi$. However the situation with $r\to \infty$ depends on the sign of $M-qQ$. If $M-qQ>0$, the arrow direction is to the right and the topological number $W=0$. Otherwise, the arrow points to left and $W=-1$. The full boundaries include four parts $I_1$, $I_2$, $I_3$, and $I_4$ corresponding to $r=\infty$, $\theta=\pi$, $r=r_h$, and $\theta=0$, respectively.}
    \label{Bca}
\end{figure}

In summary, at $r=r_h$, $\theta=0$, and $\theta=\pi$, the vector $\vec{\phi}_1=(\phi^r_1,\phi^\theta_1)$ points rightward, downward, and upward, respectively. However, the direction of the vector at $r=\infty$ is not unique and depends on the sign of $M-qQ$ as indicated in Eq. (\ref{infB}). Specifically, if $M-qQ$ is positive, the vector points to the right, resulting in a global topological number (winding number) of $W=0$. Instead, if $M-qQ<0$, the vector points to the left, yielding a topological number $W=-1$. A schematic diagram illustrating the vector constructed from $E_1$ at the boundary $\partial \Sigma$ is depicted in Fig. \ref{Bca}. It is worth pointing out that the topological number $W$ counts the number of the loops that the vector makes in the vector space when the one moves along the closed boundary $\partial\Sigma$ in a counterclockwise direction.

\begin{figure}[htb]
    \centering
    \includegraphics[width=7cm]{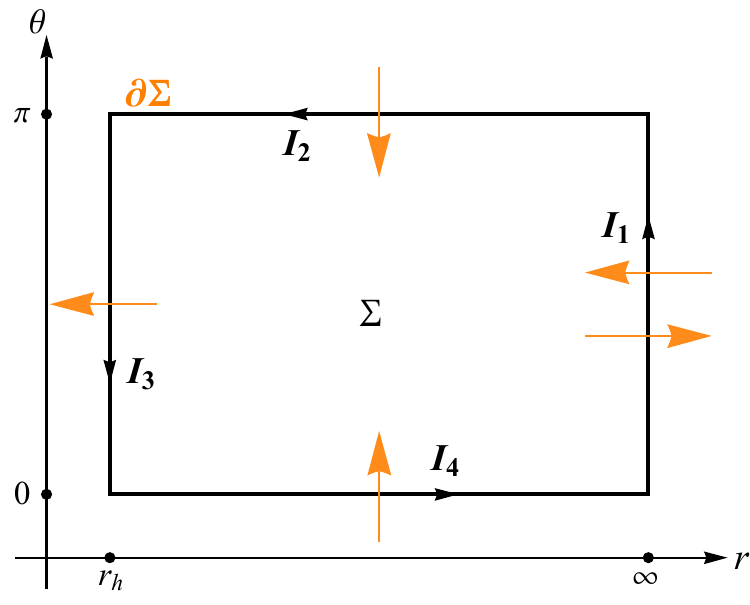}
    \caption{The boundary behaviors of vector field $\vec{\phi}_2=(\phi^r_2,\;\phi^\theta_2)$ constructed from $E_2$ on ($r,\,\theta$) plane. The orange arrows represent the direction of vector field, which are uniquely determined at $r\to r_h$ and $\theta\to 0$ or $\pi$. However, the situation with $r\to \infty$ depends on the sign of $-M-qQ$. If $-M-qQ>0$, the arrow direction is to the right and topological number $W=-1$. Otherwise, the arrow points to left and $W=0$. The full boundary includes four parts $I_1$, $I_2$, $I_3$, and $I_4$ corresponding to $r=\infty$, $\theta=\pi$, $r=r_h$, and $\theta=0$, respectively.}
    \label{BCb}
\end{figure}

It is important to note that our previous discussions focus only on the vector derived from $E_1$. However, considering the expression (\ref{eqE2}), the energy $E_2$ may also take positive values under some suitable parameter conditions. Consequently, it becomes necessary to investigate the vector field $\vec{\phi}_2=(\phi^r_2,\phi^\theta_2)$ constructed by $E_2$. Similarly, we have the following $r$- and $\theta$-components of $\vec{\phi}_2$
\begin{align}
    \phi^r_2=&\; \frac{\partial_r E_2}{\sqrt{g_{rr}}} =\frac{-\sqrt{2}  \csc(\theta ) f'(r) \sqrt{2 L^2-r^2 \cos (2 \theta)+r^2}}{4 r}- \frac{qQ \sqrt{f(r)}}{r^2} +\frac{ \sqrt{2} L^2 f(r) \csc(\theta )}{ r^2 \sqrt{2 L^2-r^2 \cos (2 \theta )+r^2}}, \label{phir2} \\
    \phi^\theta_2=&\; \frac{\partial_\theta E_2}{\sqrt{g_{\theta \theta}}} =\frac{\sqrt{2f(r)} L^2 \cot (\theta ) \csc(\theta )}{r^2 \sqrt{ 2 L^2-r^2 \cos (2 \theta )+r^2}} \label{phith2}.
\end{align}
Performing the similar treatment, we can obtain the boundary behavior of the vector $\vec{\phi}_2=(\phi^r_2,\phi^\theta_2)$. The detailed results show that the direction of the vector $\vec{\phi}_2$ exhibits an opposite direction at $r= r_h$, $\theta=0$, and $\theta=\pi$ when one compares to the $E_1$ case. This is mainly caused by the additional minus sign of $\phi^\theta_2$, see Eq. (\ref{phith2}). On the other hand, the direction of the vector at $r=\infty$ depends on the sign of $-M-qQ$ instead, which is closely related to the asymptotic behavior of $E_2$ at infinity. The corresponding schematic diagram of the vector $\vec{\phi}_2=(\phi^r_2,\phi^\theta_2)$ is illustrated in Fig. \ref{BCb}.

To provide a clearer understanding of the relationship between the sign of $M-qQ$ ($-M-qQ$) for $E_1$ ($E_2$) and the global topological number $W$ of the vector $\vec{\phi}_1$ ($\vec{\phi}_2$), we summarize the key findings in Table \ref{tab1}. In this table, we categorize the charge-to-mass ratio and the black hole charge, $qQ$, into four distinct cases:
\begin{eqnarray} \label{Clas}
    &&\text{(1) unlike strong charge:}\;qQ<-M<0; \nonumber \\
    &&\text{(2) unlike weak charge:}\;-M<qQ<0; \nonumber \\
    &&\text{(3) like weak charge:}\;0<qQ<M; \nonumber \\
    &&\text{(4) like strong charge:}\;0<M<qQ.
\end{eqnarray}
It is worth emphasizing that these four regimes correspond to distinct configurations of the TCO related to energy or angular momentum, as will be further discussed in Section \ref{sec4}.

\begin{table}[H]
\centering
\caption{The sign of the radial component of $\vec{\phi}_1$ and $\vec{\phi}_2$ at the spatial infinity and the corresponding topological number $W$ within the distinct $qQ$ regimes. Here $\pm$ denote $\phi^r>0$ or $\phi^r<0$, and the number 0 and -1 are the topological number. $E_1:\mathrm{Sign}(M-qQ)$ signifies that $\phi^r_1$ constructed from $E_1$ and its sign relies on $M-qQ$. $E_2:\mathrm{Sign}(-M-qQ)$ has analogous meaning. $W_{tot}$ is defined as the sum of $W_1$ and $W_2$ for $E_1$ and $E_2$.} \label{tab1}
\vspace{3mm}
\begin{tabular}{ccccc}
\hline
& $qQ<-M<0$ \;\;  & $-M<qQ<0$ \;\; & $0<qQ<M$ \;\; & $0<M<qQ$ \\ \hline
$E_1: \mathrm{Sign}(M-qQ)$ & + & + & + & -  \\
$W_1$  & 0 & 0 & 0 & -1 \\\hline
$E_2: \mathrm{Sign}(-M-qQ)$ & + & -& - & -\\
$W_2$ & -1 & 0  & 0 & 0\\ \hline
$W_{tot}:=W_1+W_2$& -1  & 0   & 0 & -1\\ \hline
\end{tabular}
\end{table}

In Table \ref{tab1}, the influence of strong and weak charges on the topological number is clearly illustrated by combining the results obtained from $E_1$ and $E_2$. We denote $W_{tot}$ as total topological number, and $W_1$ or $W_2$ are responsible for vector field $\vec{\phi}_1$ and $\vec{\phi}_2$ constructed from $E_1$ or $E_2$, respectively. It is evident that in the case of weak charges  ($0<|qQ|<M$), the topological number vanishes regardless of whether the charges of the particle and black hole are like or unlike. This result is consistent with the findings for the uncharged particles discussed in previous reference \cite{Ye2023gmk}, strongly indicating that the TCOs always come in pairs. However, in the case of strong charges ($M<|qQ|$), the topological number changes to -1, suggesting the existence of one more unstable TCO to the stable ones. This novel pattern in the topology of the TCO is noteworthy.

Now, we will provide a brief introduction to the concept of topological charge and present some essential conclusions. For a detailed, we would like to refer readers to Ref. \cite{Duan1984ws}. In principle, to determine the topological charge $W$ associated with a particular zero point of the vector, we need to perform the integral over a considered parameter region $\Sigma$ that only contains that zero point
\begin{align}
    W=\int_{\Sigma}j^0\; d^2x = \frac{1}{2\pi}\oint_{\partial \Sigma} \epsilon_{ab}\; n^a\; d n^b,
\end{align}
where $j^0$ is the zero-component of the topological current defined as \cite{Duan1984ws}
\begin{equation}
    j^\mu := \frac{1}{2 \pi} \epsilon^{\mu \nu \rho} \epsilon_{ab}\frac{\partial n^a}{\partial x^\nu} \frac{\partial n^b}{\partial x^\rho}.
\end{equation}
Here $x^\mu=(t,\,r,\,\theta)$ and the Latin indexes $a$, $b$ take values $r$ and $\theta$. The corresponding unit vector is $\vec{n}=(\phi^r,\phi^\theta)/\sqrt{(\phi^r)^2+(\phi^\theta)^2}$. In particular, the topological current is conserved.

In this study, we encountered three distinct types of topological configurations near the zero points. We summarize several key conclusions below:
\begin{enumerate}
    \item[$\clubsuit$] $W=0$. The vector flows outward from its zero point in the $\theta$-direction and maintains a consistent orientation (either pointing to the left or right) along the $r$-direction. This scenario corresponds to either the absence of the TCOs or the presence of the ISCOs acting as the bifurcation points.
    \item[$\spadesuit$] $W=+1$. The vector flows outward from the zero point in both the $\theta$-direction and the $r$-direction. This configuration of the zero point signifies the presence of the stable TCOs.
    \item[$\blacklozenge$] $W=-1$. The vector flows outward from the zero point in $\theta$-direction, and inward at the $r$-direction. This configuration signifies the unstable TCOs.
\end{enumerate}

\section{Reissner-Nordstr\"om black hole}\label{sec3}

In the previous section, we have studied the topology of the TCOs for the charged particles within a spherically symmetrical black hole spacetime characterized by the electromagnetic potential (\ref{Afp}). In the following sections, we will apply the general method outlined in Section \ref{sec2} to a specific background, namely the four-dimensional Reissner-Nordstr\"om (RN) black hole, which represents a static spherically symmetrical solution of Einstein field equation coupling with the Maxwell field minimally.

The line element of the RN black hole can also be described by (\ref{LE}), and the metric function is given by
\begin{equation}
    f(r)=1-\frac{2M}{r}+\frac{Q^2}{r^2}.
\end{equation}
The electromagnetic potential $A_\mu$ of Maxwell field takes the same form with Eq. (\ref{Afp}).

A brief calculation shows that the effective potential governing the motion of the charged test particles reads
\begin{equation}
    V_{eff}=\frac{-1}{1-\frac{2M}{r}+\frac{Q^2}{r^2}}(E-E_1)(E-E_2),
\end{equation}
where
\begin{align}
    &E_1=\frac{2 q Q r+\sqrt{2} \csc (\theta ) \sqrt{\left(r^2-2rM+Q^2\right) \left(2 L^2-r^2 \cos (2 \theta )+r^2\right)}}{2 r^2}, \\
    &E_2=\frac{2 q Q r-\sqrt{2} \csc (\theta ) \sqrt{\left(r^2-2rM+Q^2\right) \left(2 L^2-r^2 \cos (2 \theta )+r^2\right)}}{2 r^2}.
\end{align}
Via Eqs. \eqref{phir1} and \eqref{phith1}, the radial and angular components of the vector $\vec{\phi}_1$ associated with $E_1$ are in the following form
\begin{align}
    \phi^r_1=&\frac{\csc (\theta ) \left(2 L^2 \left(3 M r-2 Q^2-r^2\right)+r^2\left(1-\cos(2 \theta)\right) \left(M r-Q^2\right)\right)}{\sqrt{2} r^4 \sqrt{2 L^2-r^2 \cos (2 \theta)+r^2}}-\frac{q Q \sqrt{r (r-2 M)+Q^2}}{r^3},\\
    \phi^\theta_1=&-\frac{\sqrt{2} L^2 \cot (\theta ) \csc(\theta )\sqrt{r^2-2Mr+Q^2}}{r^3 \sqrt{2 L^2-r^2 \cos (2 \theta )+r^2}}.
\end{align}
Using the similar manner, the $\vec{\phi}_2$ relating to $E_2$ can also be obtained from the \eqref{phir2} and \eqref{phith2}
\begin{align}
    \phi^r_2=&\frac{-\csc (\theta ) \left( 2 L^2 \left(3 M r-2 Q^2-r^2\right)+r^2 (1-\cos (2 \theta ) ) (M r-Q^2)\right)}{\sqrt{2} r^4 \sqrt{2 L^2-r^2 \cos (2 \theta )+r^2}}-\frac{q Q \sqrt{r (r-2 M)+Q^2}}{r^3},\\
    \phi^\theta_2=&\frac{\sqrt{2} L^2 \cot (\theta ) \csc (\theta )\sqrt{r^2-2rM+Q^2}}{r^3 \sqrt{2 L^2-r^2 \cos (2 \theta )+r^2}}.
\end{align}

The zero points of the vector $\vec{\phi}_1$ and $\vec{\phi}_2$ correspond to the locations of the TCOs. By solving these zero points, we can easily determine the energy and angular momentum of the TCOs
\begin{align}
    &E_{t1}=\frac{qQr (r-4 M)+3q Q^3+\left(r^2-2r M+Q^2\right) \sqrt{4 r (r-3 M)+\left(q^2+8\right) Q^2}}{2 r \left(r^2-3rM+2 Q^2\right)},\label{EE1}\\
    &L_{t1}= r\sqrt{\tfrac{Q^2 r \left(\left(q^2-2\right) r-2 M \left(q^2-5\right)\right)-q Q \left(r^2-2rM+Q^2\right) \sqrt{4 r (r-3 M)+\left(q^2+8\right) Q^2}+2 M r^2 (r-3 M)+\left(q^2-4\right) Q^4}{2\left(r^2-3rM+2 Q^2\right)^2}};
\end{align}
\begin{align}
    &E_{t2}=\frac{qQr (r-4 M)+3q Q^3-\left(r^2-2rM+Q^2\right) \sqrt{4 r (r-3 M)+\left(q^2+8\right) Q^2}}{2 r \left(r^2-3rM+2 Q^2\right)},\\
    &L_{t2}=r \sqrt{\tfrac{Q^2 r \left(\left(q^2-2\right) r-2 M \left(q^2-5\right)\right)+q Q \left(r^2-2rM+Q^2\right) \sqrt{4 r (r-3 M)+\left(q^2+8\right) Q^2}+2 M r^2 (r-3 M)+\left(q^2-4\right) Q^4}{2\left(r^2-3rM+2 Q^2\right)^2}}\label{LL2}.
\end{align}
From above obtained expressions for the energy and angular momentum, we have three specific radii as follows
\begin{align}
    &r_L=\frac{3M+\sqrt{9M^2-8Q^2-q^2Q^2}}{2}, \label{rL}\\
    &r_P= \frac{3M+\sqrt{9M^2-8Q^2}}{2}, \label{rP}\\
    &r_S=\frac{MQ^2(q^2-1)+Q^2 \sqrt{q^2(q^2-1)(M^2-Q^2)}}{q^2Q^2-M^2} \label{rS}.
\end{align}
For $r<r_L$, both energies $E_{t1}$ and $E_{t2}$ become imaginary. Therefore, we only restrict our attention to $r>r_L$. The radius $r_P$ corresponds to the case where $r^2-3rM+2Q^2=0$, resulting in divergences of $E_{t1}$, $L_{t1}$, $E_{t2}$, and $L_{t2}$. On the other hand, the radius $r_S$ indicates situations where either $L_{t1}=0$ or $L_{t2}=0$, signifying the absence of angular momentum.

Using \eqref{IS1} and \eqref{IS2}, the conditions determined the radius of the ISCO, i.e., $r_{ISCO}(E_{t1},L_{t1})$ and $r_{ISCO}(E_{t2},L_{t2})$, reduce to
\begin{eqnarray}
    &Q^2 r^2 \left(6 M^2 \left(q^2+13\right)-2 M \left(2 q^2+11\right) r+q^2 r^2\right)+Q^4 r \left(\left(q^2+8\right) r-6 M \left(q^2+10\right)\right)-2 M r^3 \left(18 M^2-9 M r+r^2\right) \nonumber\\
    &+q Q \left(r^2 \left(12 M^2-8 M r+r^2\right)+Q^2 r (7 r-18 M)+6 Q^4\right) \sqrt{4 r (r-3 M)+\left(q^2+8\right) Q^2}+2 \left(q^2+8\right) Q^6=0, \label{RNISa}\\
    &Q^2 r^2 \left(6 M^2 \left(q^2+13\right)-2 M \left(2 q^2+11\right) r+q^2 r^2\right)+Q^4 r \left(\left(q^2+8\right) r-6 M \left(q^2+10\right)\right)    -2 M r^3 \left(18 M^2-9 M r+r^2\right) \nonumber\\
    &-q Q \left(r^2 \left(12 M^2-8 M r+r^2\right)+Q^2 r (7 r-18 M)+6 Q^4\right) \sqrt{4 r (r-3 M)+\left(q^2+8\right) Q^2}+2 \left(q^2+8\right) Q^6=0. \label{RNISb}
\end{eqnarray}
These two equations demonstrate that the ISCO is determined by the black hole mass $M$ and charge $Q$, as well as the charge-to-mass ratio $q$ of the charged particle. In the case of neutral particles, where we take the limit $q\to 0$, these two equations become dependent on each other. However, in the more general scenario where $q\neq 0$, one can solve $r_{ISCO}(E_{t1}, L_{t1})$ and $r_{ISCO}(E_{t2}, L_{t2})$. Nevertheless, it is important to note that one of these solutions yields a negative energy, as discussed in Section \ref{sec4}.

\section{Topological configurations of timelike circular orbits} \label{sec4}

In this section, we investigate the topological configurations of the TCOs in four distinct regimes based on the $qQ$ classification (\ref{Clas}). These regimes are as follows: the unlike strong charge regime ($qQ<-M<0$), the unlike weak charge regime  ($-M<qQ<0$), the like weak charge regime ($0<qQ<M$), and the like strong charge regime ($0<M<qQ$). For each regime, we will determine the topological configuration of the zero point of the vector $\vec{\phi}_1$ and $\vec{\phi}_2$ against the $E_1$ and $E_2$ on the ($r,\theta$) plane, respectively. We will then analyze the energy and angular momentum of the TCOs. Finally, we will discuss the representation of TCO radius and the topological charge as a function of angular momentum.

\subsection{Unlike Strong Charge Regime: $qQ<-M<0$}

We begin by examining the regime $qQ<-M<0$, where the electric charge type of the charged particle is opposite to that of the charged black hole. Throughout our analysis, we focus on non-extremal charged black holes, specifically $0<|Q|<M$, while fixing the parameters $Q=0.6$ and $M=1$ without loss of generality.

\begin{figure}[h]
    \centering
    \subfigure[]{\includegraphics[width=6cm]{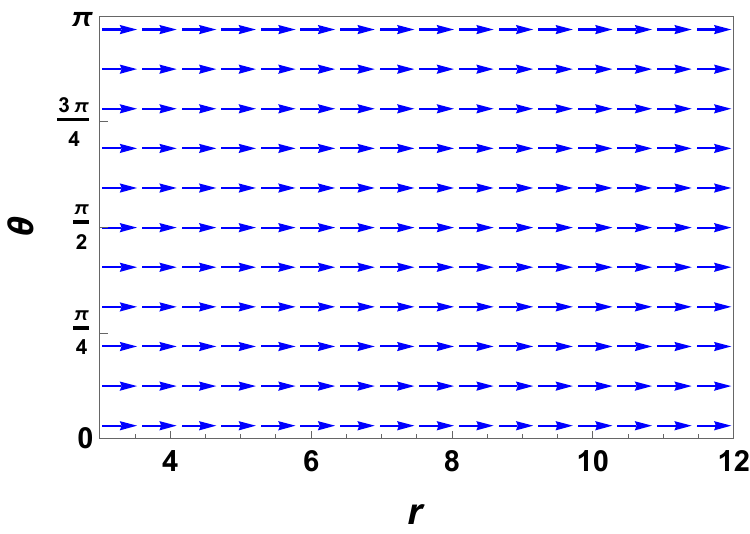}\label{O1t1}}\hspace{6mm}
    \subfigure[]{\includegraphics[width=6cm]{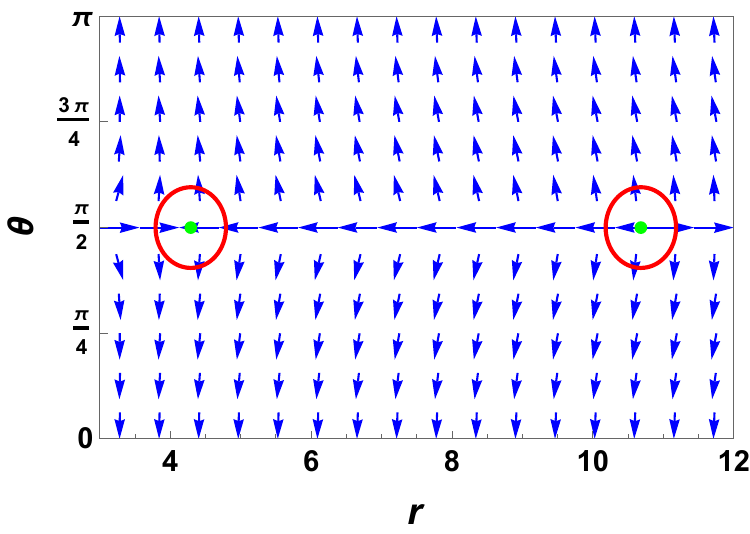}\label{O1t2}}
    \caption{Unit vector $\vec{n}_1=(\phi^r_1,\phi^\theta_1)/||\vec{\phi}_1||$ constructed from $E_1$ on ($r,\theta$) plane in the unlike strong charge regime with $q=-3$. (a) Angular momentum $L=5.5<L_{ISCO}$. (b) $L=6.5>L_{ISCO}$. Two green zero points of $\vec{n}$ locate at $r=4.2902$ and $10.6808$. Here $L_{ISCO}=5.8976$.}
    \label{O1top}
\end{figure}

For a strong charge $q=-3$, the asymptotic analysis of the vector $\vec{\phi}_1=(\phi^r_1, \phi^\theta_1)$ for $E_1$ at the boundary $\partial\Sigma$ reveals a vanishing topological charge $W_1=0$, as shown in Table \ref{tab1}. This corresponds to two distinct topological configurations of the unit vector $\vec{n}_1$ on the $(r, \theta)$ plane, which are depicted in Figs. \ref{O1t1} and \ref{O1t2}. In the first case, considering an angular momentum value of $L=5.5<L_{ISCO}=5.8976$, there are no zero points presented in the region $\Sigma$. In Fig. \ref{O1t1}, the arrows of $\vec{n}_1$ consistently point towards the right along the $r$-direction at $\theta=\pi/2$. This pattern confirms the previous conclusion $\clubsuit$, indicating the absence of the TCOs and resulting in a vanishing topological charge. On the other hand, by increasing the angular momentum to $L=6.5>L_{ISCO}$, we can clearly observe two zero points located at $r=4.2902$ and $10.6808$, and enclosed by the red closed path in Fig. \ref{O1t2}. Examining the left zero point, one can see that the vector arrows flow into it along the $r$-direction and out of it along the $\theta$-direction. Referring to the conclusion $\loze$, this zero point corresponds to an unstable TCO with topological charge $W=-1$. As for the right zero point, the vector arrows all flow out of it in both the $r$ and $\theta$ directions, which indicates that it corresponds to a stable TCO with $W=+1$. Summing these topological charges yields the topological number $W_1=-1+1=0$ as expected.

\begin{figure}[h]
    \centering
    \includegraphics[width=7cm]{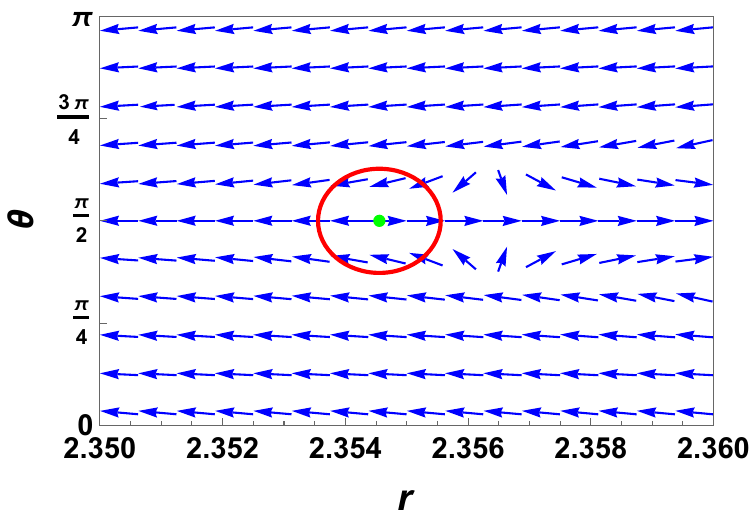}
    \caption{Unit vector $\vec{n}_2=(\phi^r_2,\phi^\theta_2)/||\vec{\phi}_2||$ constructed from $E_2$ on ($r,\theta$) plane in the unlike strong charge regime with $q=-3$. Here $L=2.2$ and the green zero point locates at $r=2.3545$. The vector configuration is similar for arbitrary positive angular momentum.}
    \label{O1t3}
\end{figure}

From the Table \ref{tab1}, we already know that the total topological number of TCO in the unlike strong charge regime is $W_{tot}=-1$. Nevertheless, the above result on vector field $\vec{\phi}_1$ built from the $E_1$ is just $W_1=0$. Thereby, we should continue to consider the situation of vector field $\vec{\phi}_2$ coming from $E_2$. We show the topological configuration of the unit vector $\vec{n}_2= \vec{\phi}_2/||\vec{\phi}_2||$ in Fig. \ref{O1t3}. Contrary to the case of $E_1$, here only one green zero point can be found at $r=2.3545$ for $\vec{n}_2$. Based on the conclusion $\loze$, this zero point represents an unstable TCO with topological number $W_2=-1$. Even thought the angular momentum here we chose is $L=2.2$, the outcome of $W_2=-1$ holds for arbitrary $L>0$. Combining $W_1$ and $W_2$ for the vector field $\vec{\phi}_1$ and $\vec{\phi}_2$, we end up with the total topological number $W_{tot}=W_1+W_2=-1$ in the unlike strong charge regime. This strongly verifies our analysis in Section \ref{sec2}.

\begin{figure}[htb]
    \centering
    \subfigure[]{\includegraphics[width=6cm]{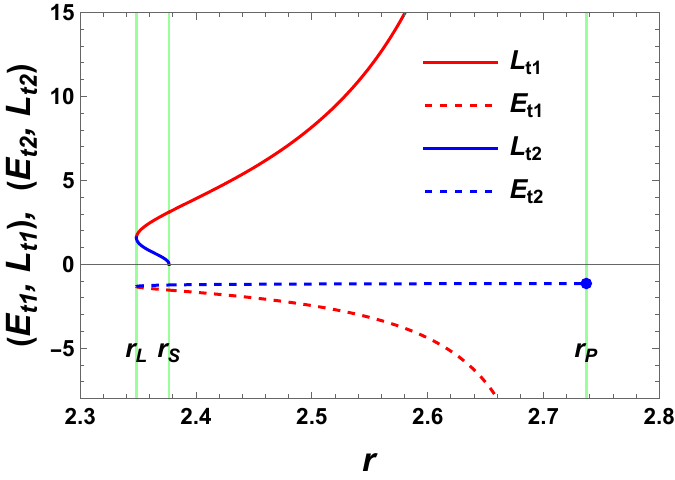}\label{O1ELa}} \hspace{6mm}
    \subfigure[]{\includegraphics[width=6cm]{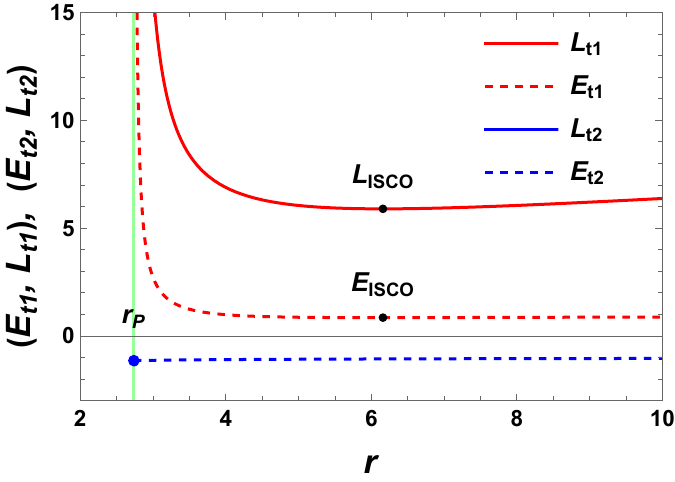}\label{O1ELb}}
    \caption{The solutions $(E_{t1},L_{t1})$ and $(E_{t2},L_{t2})$ of TCOs in the unlike strong charge regime with $q=-3$. (a) Interval [$r_L,r_P$]. (b) Interval [$r_P,\infty$). The blue point indicates that this curve can be extended and surpass $r_P$. The location of ISCO is denoted as the black point. Here $r_L=2.3485$, $r_P=2.7369$, and $r_S=2.3766$.}
\end{figure}

Now, let us proceed with the analysis of the energy and angular momentum of the TCOs. As expected, we have two sets of solutions, namely $(E_{t1},L_{t1})$ and $(E_{t2},L_{t2})$ given in Eqs. (\ref{EE1})-(\ref{LL2}). It is worth noting that we split the radial range of TCO into two parts, i.e., $[r_L,r_P]$ and $[r_P, \infty)$ using the \eqref{rL} and \eqref{rP}. The solutions $(E_{t1},L_{t1})$ and $(E_{t2},L_{t2})$ in these two intervals are separately presented in Figs. \ref{O1ELa} and \ref{O1ELb}. In Fig. \ref{O1ELa}, the red and blue solid curves denoting $L_{t1}$ and $L_{t2}$ are connected at $r_L$, while $L_{t1}$ is always increasing from $r_L$ to $r_P$ and the value of $L_{t2}$ eventually decays to zero at $r_S$. Hence $L_{t1}$ and $L_{t2}$ can form a sole TCO branch in $L\in [0,\infty)$ within the interval $[r_L,r_P)$. If we focus on $E_{t1}$ and $E_{t2}$ described by red and blue dashed curves, we observe that they are both negative and exist in all $[r_L,r_P]$. Different from the decreasing $E_{t1}$, we notice that $E_{t2}$ can extend $r_P$ to $r=\infty$, so we label the position of $E_{t2}$ at $r_P$ with a blue point, indicating this curve can exceed it. In Fig. \ref{O1ELb}, the new branch of $L_{t1}$ and $E_{t1}$ appears in the range $[r_P,\infty)$. Both of them are positive, and the curves firstly descend with $r$ and then slowly grow. The minimal point of the curves corresponds to ISCO, and its location, energy and angular momentum can be solved from the Eq. \eqref{RNISa}
\begin{align}
    &r_{ISCO}=6.1593, \quad E_{t1}=0.853, \quad L_{t1}=5.8976.
\end{align}
It is noteworthy that $L_{t1}=5.8976$ represents a critical value distinguishing two distinct topological configurations of the unit vector $\vec{n}_1$ depicted in Figs. \ref{O1t1} and \ref{O1t2}. Despite the negative $E_{t2}$ can emerge in Fig. \ref{O1ELb}, it not provides us extra information on TCO because of the absence of $L_{t2}$.

One important fact is that the combined branch  $L_{t1}\cup L_{t2}$ and the branch $L_{t1}$ with ISCO in Figs. \ref{O1ELa} and \ref{O1ELb} can separately interpret the topological number $W_1=0$ and $W_2=-1$ of TCO  given by Figs. \ref{O1top} and \ref{O1t3}.

\begin{figure}[htb]
    \centering
    \includegraphics[width=7cm]{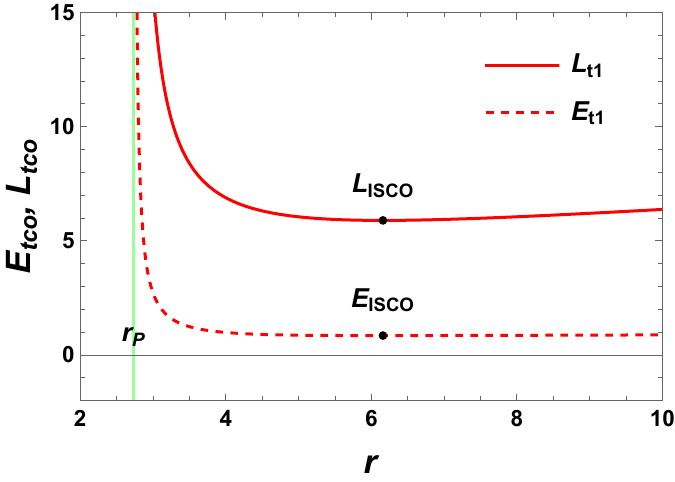}
    \caption{Valid energy $E_{tco}$ and angular momentum $L_{tco}$ of TCO in the unlike strong charge regime with $q=-3$. The location of ISCO is denoted as the black point. Here $r_P=2.7369$.}
    \label{O1ELc}
\end{figure}

For RN spacetime, if we focus exclusively on the positive energy values of TCO, then by disregarding the negative energy portion, we obtain the valid energy $E_{tco}$ and angular momentum $L_{tco}$, which can be expressed as
\begin{align}
    &E_{tco}(r)=\left\{E_{t1}(r_a) \bigcup E_{t2}(r_b)  \,\bigg|\, r_a\in (r_P,\infty), r_b=\varnothing, qQ<-M<0 \right\},  \\
    &L_{tco}(r)=\left\{L_{t1}(r_a) \bigcup L_{t2}(r_b) \,\bigg|\, r_a\in (r_P,\infty), r_b=\varnothing, qQ<-M<0 \right\}.
\end{align}
We present them in Fig. \ref{O1ELc}. It is easy to see that $E_{t2}$ and $L_{t2}$ do not contribute to $E_{tco}$ and $L_{tco}$. The contribution only arises from the segment of $L_{t1}$ and $E_{t1}$ with $r\in (r_P,\infty)$.

\begin{figure}
    \centering
    \subfigure[]{\includegraphics[width=6cm]{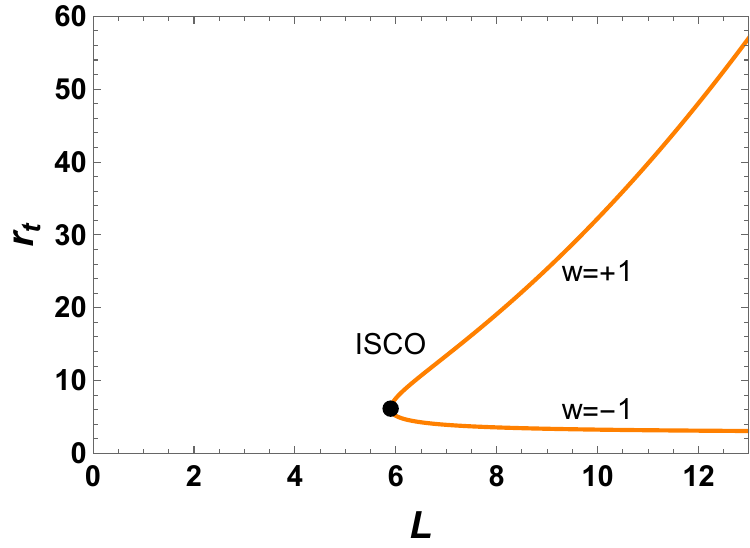}\label{O1rL}}\hspace{6mm}
    \subfigure[]{\includegraphics[width=6cm]{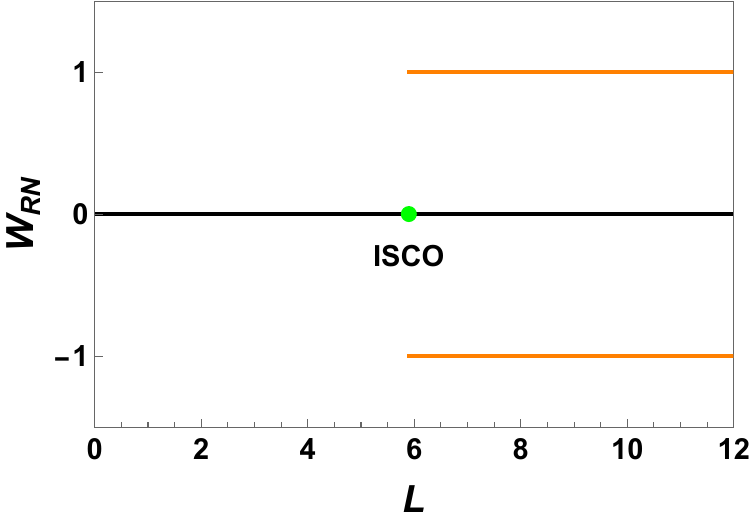}\label{O1wL}}
    \caption{(a) TCO radius $r_t$ as a function of angular momentum $L$.  (b) Topological number $W_{RN}$ as a function of $L$. Here $W=\pm 1$ represent stable and unstable TCOs. The black point denotes ISCO that is a generated bifurcation point. The orange line stands for TCO branch.}
\end{figure}

With regard to the valid energy and angular momentum, we further present the TCO radius $r_t$ as a function of the angular momentum $L$ in Fig. \ref{O1rL}. It becomes evident that when $L<L_{ISCO}$, no TCO can emerge. However, if $L>L_{ISCO}$, both stable and unstable TCO branches originate from the ISCO. This observation leads us to consider the ISCO as a bifurcation point \cite{Fu2000pb}. Moreover, in Fig. \ref{O1wL}, we depict the topological charge of the TCO branches with orange lines and the total topological number with black line. The stable TCOs possess a topological charge of $W=+1$, while the unstable TCOs have a topological charge of $W=-1$. Obviously, when $L>L_{ISCO}$, both the stable and unstable TCOs branches are generated simultaneously, thus the total topological number is given by $W_{RN}=+1-1=0$ when we neglect the negative energy sector. Regardless of the value of the angular momentum, the total topological number always vanishes, as illustrated by the black line in figure. This suggests that the valid topological number $W_{RN}=0$ for RN black hole is significant distinguished from $W_{tot}=-1$ in Table \ref{tab1}.

\subsection{Unlike Weak Charge Regime: $-M<qQ<0$}

For this case, we would like to take the charge-to-mass ratio $q=-0.5$ as an example. Due to the weak charge $q$, the Coulomb attractive force is expected to be weaker than the gravitational interaction of the RN black holes.

\begin{figure}[hbt]
    \centering
    \subfigure[]{\includegraphics[width=6cm]{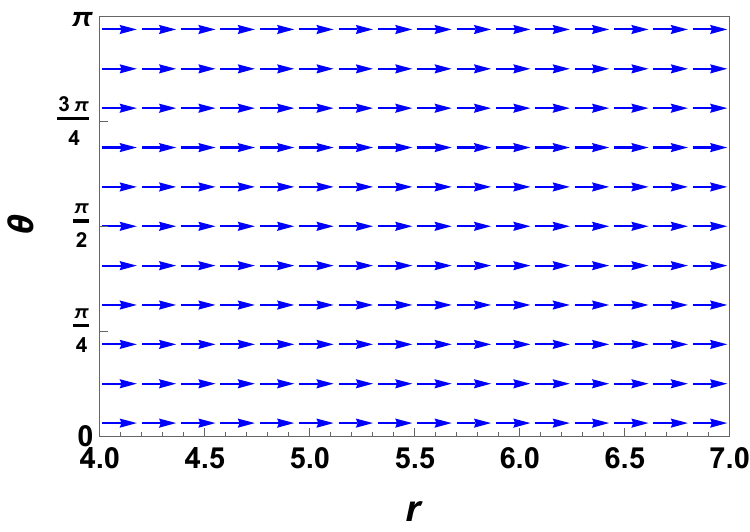}\label{O2t1}} \hspace{6mm}
    \subfigure[]{\includegraphics[width=6cm]{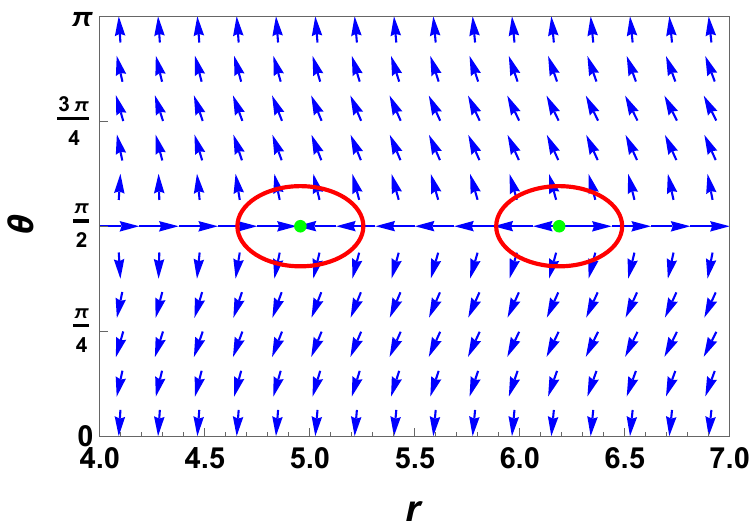}\label{O2t2}}
    \caption{Unit vector $\vec{n}_1=(\phi^r_1,\phi^\theta_1)/||\vec{\phi}_1||$ constructed from $E_1$ on ($r,\,\theta$) plane in the unlike weak charge regime with $q=-0.5$. (a) Angular momentum $L=3.73<L_{ISCO}$. (b) $L=3.78>L_{ISCO}$. Two green zero points locate at $r=4.9549$ and $6.1876$. Here $L_{ISCO}=3.7561$.}
    \label{O2top1}
\end{figure}

According to the Table \ref{tab1}, the boundary behavior of the vector field $\vec{\phi}_1$ for $E_1$ indicates that the topological number in the region $\Sigma$ is $W_1=0$, which is consistent with the presence of the two distinct topological configurations of the unit vector $\vec{n}_1$ depicted in Figs. \ref{O2t1} and \ref{O2t2}. The first configuration, characterized by an angular momentum of $L=3.73$, does not possess a zero point. In this case, the direction of the vector always points towards the right at $\theta=\pi/2$ with $r$. Utilizing the conclusion $\clubsuit$, we can easily obtain the result that the topological charge $W=0$ for this particular configuration.

By increasing the angular momentum, such as $L=3.78$, two zero points of the unit vector $\vec{n}_1$, marked in green color, emerge at $r=4.9549$ and $6.1876$. The vector arrows near the left zero point flow into it along the $r$-direction and flow out of it along the $\theta$-direction. This behavior is consistent with the conclusion $\loze$, and gives a topological charge $W=-1$ for this zero point. As for the right zero point, all arrows flow out from it in both $r$ and $\theta$ directions, resulting in a topological charge  $W=+1$. Combining with these results, we have the topological number $W=-1+1=0$ for the second configuration. Consequently, the results obtained for these two configurations are consistent with $W_1=0$ for $E_1$ presented in Table \ref{tab1}.

\begin{figure}[htb]
    \centering
    \subfigure[]{\includegraphics[width=6cm]{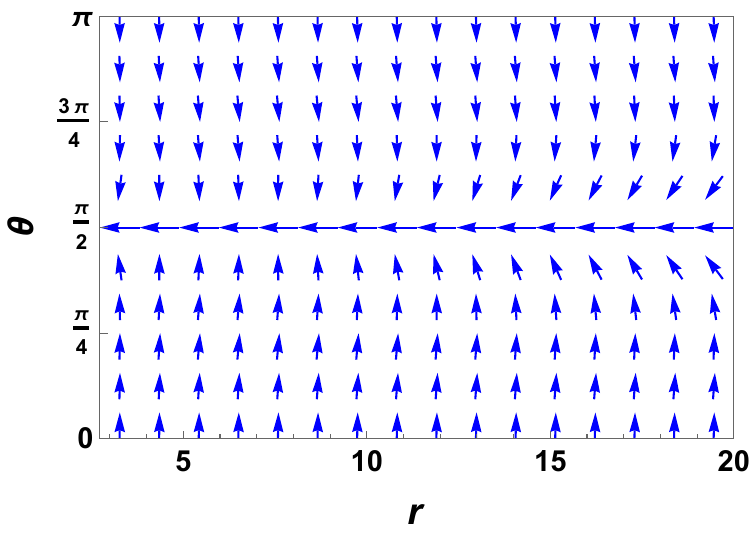}\label{O2t3}} \hspace{6mm}
    \subfigure[]{\includegraphics[width=6cm]{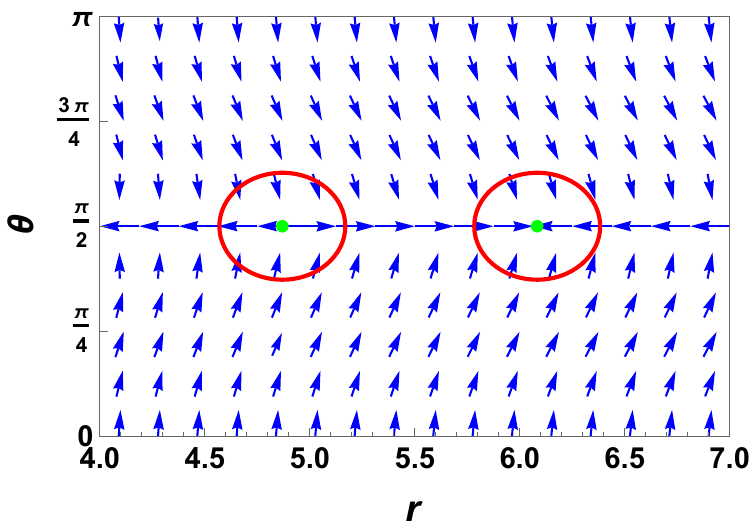}\label{O2t4}}
    \caption{Unit vector $\vec{n}_2=(\phi^r_2,\phi^\theta_2)/||\vec{\phi}_2||$ constructed from $E_2$ on ($r,\,\theta$) plane in the unlike weak charge regime with $q=-0.5$. (a) Angular momentum $L=2.75<L_{ISCO}$. (b) $L=2.78>L_{ISCO}$. Two green zero points locate at $r=4.969$ and 6.083. Here $L_{ISCO}=2.7631$}
    \label{O2top2}
\end{figure}

After investigating the vector field $\vec{\phi}_1$ built from $E_1$, we continue to explore the scenario of vector $\vec{\phi}_2$ for $E_2$. In this case, the topological number based on the Table \ref{tab1} is $W_2=0$ for $E_2$ as same as $W_1$. This outcome corresponds to two kinds of topological configuration of $\vec{n}_2$, and we exhibit them in Figs. \ref{O2t3} and \ref{O2t4}, respectively. It is quite obvious that when $L<L_{ISCO}$ there is no zero points, whereas for $L>L_{ISCO}$ two green zero points separately appear at $r=4.969$ and 6.083 with topological number $W=-1$ and +1 by applying the previous conclusions $\loze$  and $\spad$. Therefore, the topological number $W_2=0$ always holds for these two distinct configurations. Regardless of vector field constructed from $E_1$ and $E_2$, the result of vanishing topological number leads to total topological number $W_{tot}=W_1+W_2=0$ in the unlike weak charge regime as predicted in Table \ref{tab1}.

\begin{figure}
    \centering
    \subfigure[]{\includegraphics[width=6cm]{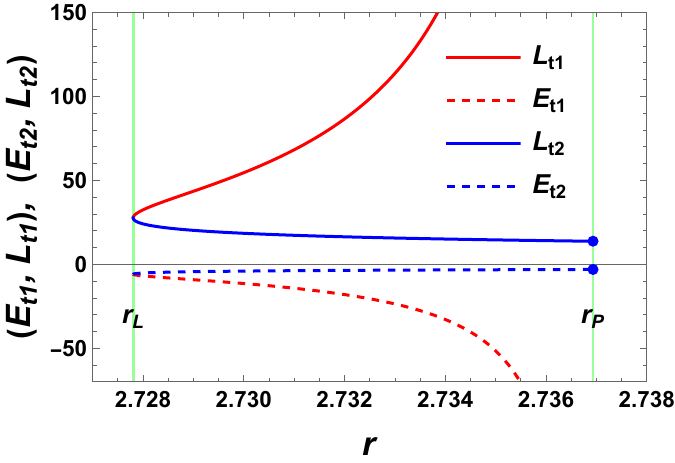}\label{O2ELa}} \hspace{6mm}
    \subfigure[]{\includegraphics[width=6cm]{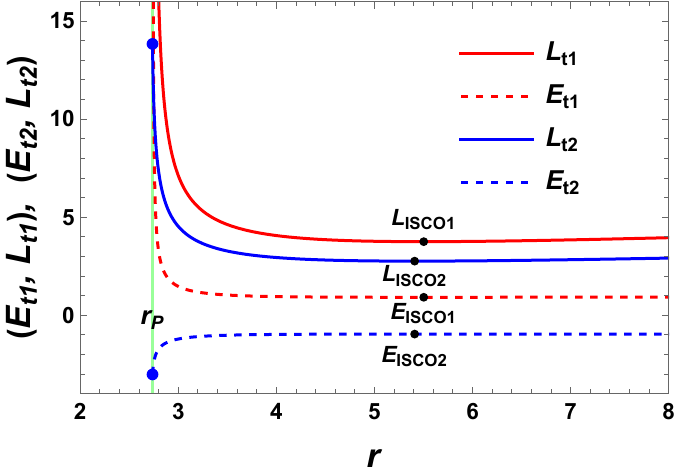}\label{O2ELb}}
    \caption{The solutions $(E_{t1},L_{t1})$ and $(E_{t2},L_{t2})$ of TCOs in the unlike weak charge regime with $q=-0.5$ (a) Interval [$r_L,r_P$]. (b) Interval [$r_P,\infty$). The blue point indicates that this curve can be extended and surpass $r_P$. The location of ISCO is denoted as the black point. Here $r_L=2.7278$ and $r_P=2.7369$.}
\end{figure}

Furthermore, we investigate the energy and angular momentum of the TCOs within the regime of weak unlike charge. Two sets of the solutions $(E_{t1}, L_{t1})$ and $(E_{t2}, L_{t2})$ situating at the intervals $[r_L,r_P]$ and $[r_P,\infty)$ are clearly presented in Figs. \ref{O2ELa} and \ref{O2ELb}, respectively. In Fig. \ref{O2ELa}, we see that $E_{t1}$ and $L_{t1}$ both start at $r_L$ and terminate at $r_P$, while they show diverse trends, namely $L_{t1}>0$ keeps increase and $E_{t1}<0$ decays within this interval. Notably, positive $L_{t2}$ and negative $E_{t2}$ at starting point $r_L$ connect with $L_{t1}$ and $E_{t1}$, respectively. Furthermore, these two blue solid and dashed curves can surpass the restriction of $r_P$ and extend to $[r_P,\infty)$ showed in Fig. \ref{O2ELb}. The blue points at $r_P$ mean that the corresponding curves can continue to be extended. Let us turn our attention to the Fig. \ref{O2ELb}. There are new $L_{t1}$ and $E_{t1}$ branches presented in $[r_P,\infty)$. Both of them have positive values and originally reduce and then gradually rise with the TCO radius $r$. Additionally, the extremal points of $L_{t1}$ and $E_{t1}$ stand for an ISCO. On the other hand, the segments of $E_{t2}$ and $L_{t2}$ locating in [$r_P,\infty$) also possess another ISCO. Through solving the constraint relations \eqref{RNISa} and \eqref{RNISb}, the radial radius, energy, and angular momentum of these two ISCOs can be given by
\begin{align}
    &r_{ISCO1}=5.5021, \quad E_{t1}=0.9204, \quad L_{t1}=3.7561;\\
    &r_{ISCO2}=5.4094, \quad E_{t2}=-0.9555, \quad L_{t2}=2.7631.
\end{align}

It is important to note that the branch $L_{t1}$ with $r_{ISCO1}$ and the combined branch $L_{t1}\cup L_{t2}$ with $r_{ISCO2}$ in Figs. \ref{O2ELb} and \ref{O2ELa} can separately interpret the topological number $W_1=0$ and $W_2=0$ of TCO  given by Figs. \ref{O2top1} and \ref{O2top2}.

\begin{figure}[htb]
    \centering
    \includegraphics[width=7cm]{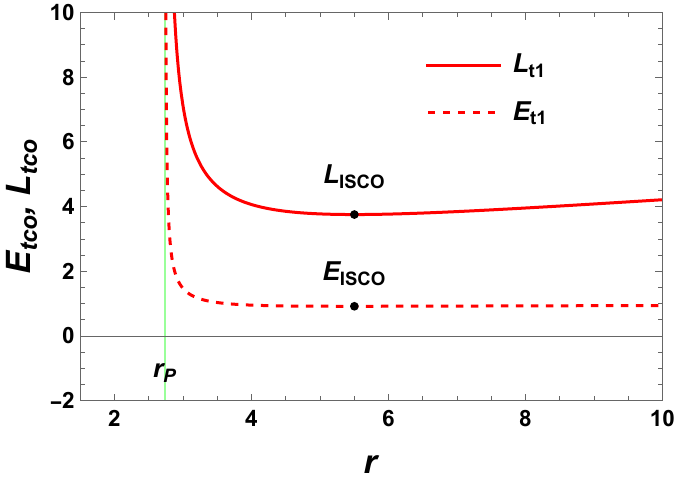}
    \caption{Valid energy $E_{tco}$ and angular momentum $L_{tco}$ of TCO in the unlike weak charge regime with $q=-3$. The location of ISCO is denoted as the black point. Here $r_P=2.7369$.}
    \label{O2ELc}
\end{figure}

Since we only focus on the TCOs with positive energy, the valid energy $E_{tco}$ and angular momentum $L_{tco}$ should be
\begin{align}
    &E_{tco}(r)=\left\{E_{t1}(r_a) \bigcup E_{t2}(r_b)  \,\bigg|\, r_a\in (r_P,\infty), r_b=\varnothing,-M<qQ<0\right\}, \\
    &L_{tco}(r)=\left\{L_{t1}(r_a) \bigcup L_{t2}(r_b) \,\bigg|\, r_a\in (r_P,\infty), r_b=\varnothing, -M<qQ<0\right\}.
\end{align}
These two expressions are shown in Fig. \ref{O2ELc} and indicates that only $E_{t1}$ and $L_{t1}$ in interval $(r_P,\infty)$ make a valid contribution to TCOs.

\begin{figure}
    \centering
    \subfigure[]{\includegraphics[width=6cm]{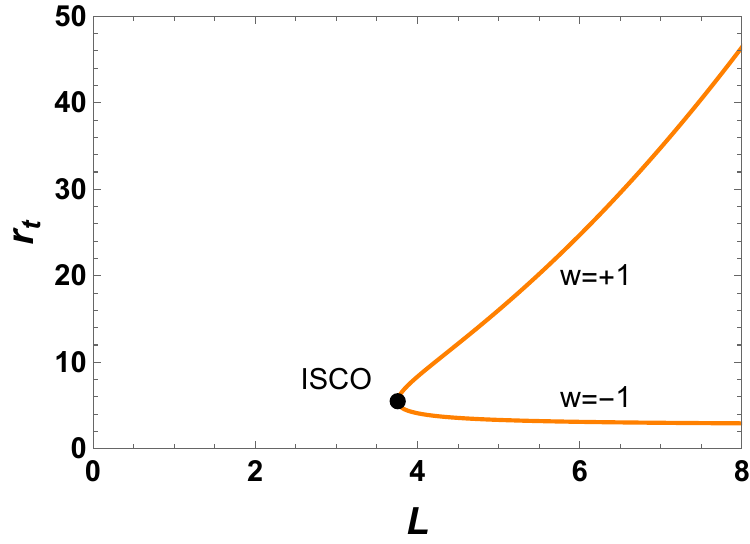}\label{O2rL}} \hspace{6mm}
    \subfigure[]{\includegraphics[width=6cm]{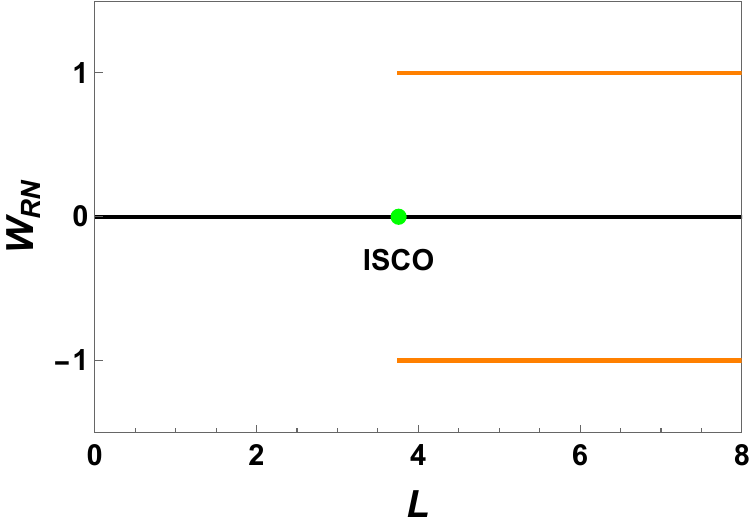}\label{O2wL}}
    \caption{(a) TCO radius $r_t$ as a function of angular momentum $L$.  (b) Topological number $W_{RN}$ as a function of $L$. Here $W=\pm 1$ represent stable and unstable TCOs. The black point denotes ISCO that is a generated bifurcation point. The orange line stands for TCO branch.}
\end{figure}

Considered the valid energy and angular momentum, we now examine the relationship between the TCO radius $r_t$ and the angular momentum $L$, as illustrated in Fig. \ref{O2rL}. The connection between the ISCO and TCO branches becomes evident. If $L<L_{ISCO}$, no TCOs can emerge. However, by increasing the angular momentum to $L>L_{ISCO}$, stable and unstable TCOs appear in pairs. These observations are consistent with these two distinct topological configurations of the vector depicted in Figs. \ref{O2t1} and \ref{O2t2}. Note that $r_{ISCO}(E_{t1},L_{t1})$, represented by the black point in the $r_{t}-L$ relation, acts as a generated bifurcation point. The topological charge of the TCO branches (depicted in orange) is shown in Fig. \ref{O2wL}. When TCO branches are absent, the topological number vanishes. In the presence of TCO branches ($L>L_{ISCO}$), since the stable and unstable TCOs possess opposite topological charges, the total topological number still vanishes. Therefore, the total topological number always remains zero, i.e., $W_{RN}=0$ within the regime of weak unlike charge, as indicated by the black line in Fig. \ref{O2wL}. This result is in accordance with $W_{tot}=0$ in the Table \ref{tab1} which do not take into account the negative energy.

\subsection{Like Weak Charge Regime: $0<qQ<M$}

As demonstrated earlier, we have studied the case of unlike charge with $qQ<0$. Now, let us turn our attention to the like charge case with $qQ>0$. To illustrate it, we consider a specific example where $q=1.2$ within the regime of like weak charge, i.e., $0<qQ<M$.

\begin{figure}[htb]
    \centering
    \subfigure[]{\includegraphics[width=6cm]{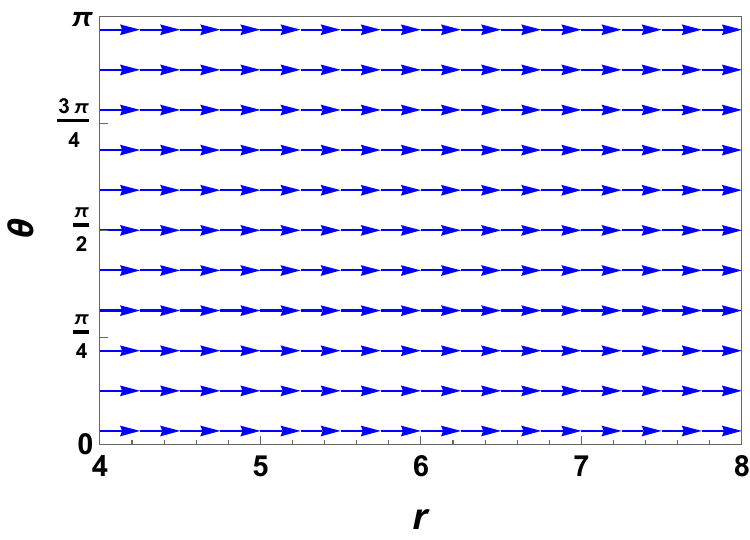}\label{S1t1}} \hspace{6mm}
    \subfigure[]{\includegraphics[width=6cm]{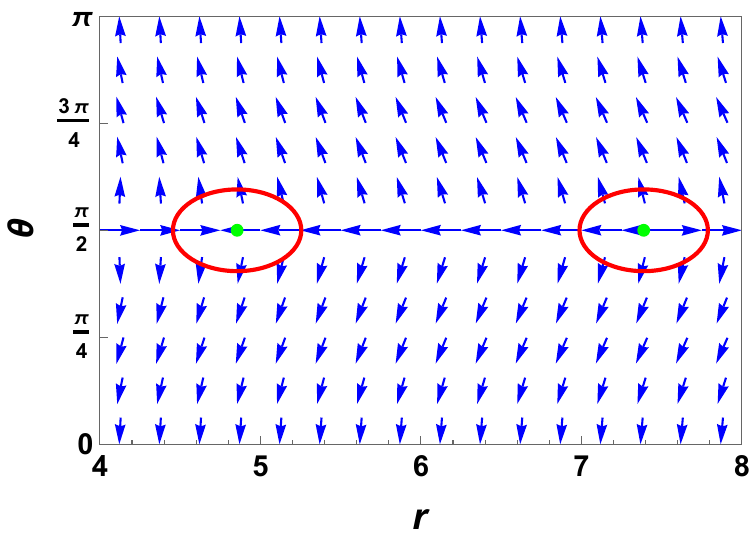}\label{S1t2}}
    \caption{Unit vector $\vec{n}_1=(\phi^r_1,\phi^\theta_1)/||\vec{\phi}_1||$ constructed from $E_1$ on ($r,\,\theta$) plane in the like weak charge regime with $q=1.2$. (a) Angular momentum $L=1.85<L_{ISCO}$. (b) $L=1.95>L_{ISCO}$. Two green zero points locate at $r=4.8549$ and 7.3882. Here $L_{ISCO}=1.9160$.}
    \label{S1top1}
\end{figure}

In this scenario, two distinct types of topological configurations of the unit vector $\vec{n}_1$ constructed from $E_1$  with the angular momentum $L=1.85$ and $1.95$ are depicted in Figs. \ref{S1t1} and \ref{S1t2}. In the former case, there is an absence of the TCOs, whereas in the latter case, unstable and stable TCOs appear in pairs at $r=4.8549$ and $7.3882$. It is noteworthy that these two configurations are similar to the previous case depicted in Figs. \ref{O1t1} and \ref{O1t2}, as well as Figs. \ref{O2t1} and \ref{O2t2}. Regardless of the occurrence of the TCOs, the topological number remains zero, i.e., $W_1=0$ for the vector of $E_1$. Furthermore, the angular momentum $L_{ISCO}$ of the ISCO can be used to split these two topological configurations.

\begin{figure}[htb]
    \centering
    \subfigure[]{\includegraphics[width=6cm]{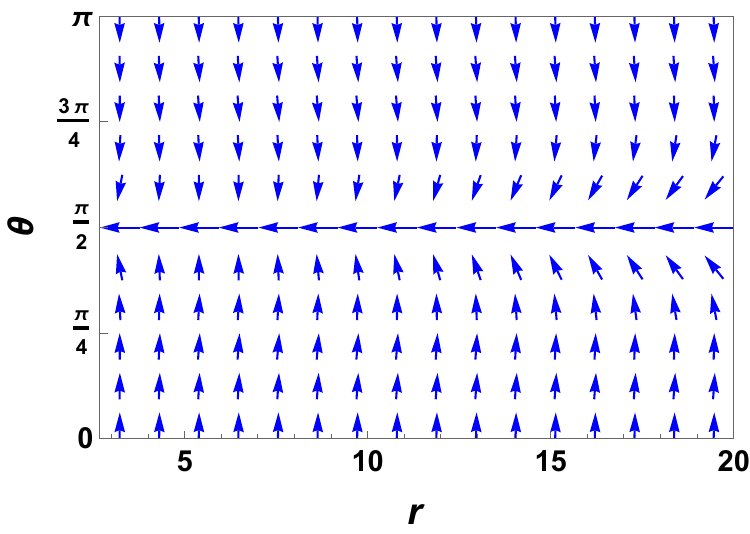}\label{S1t3}} \hspace{6mm}
    \subfigure[]{\includegraphics[width=6cm]{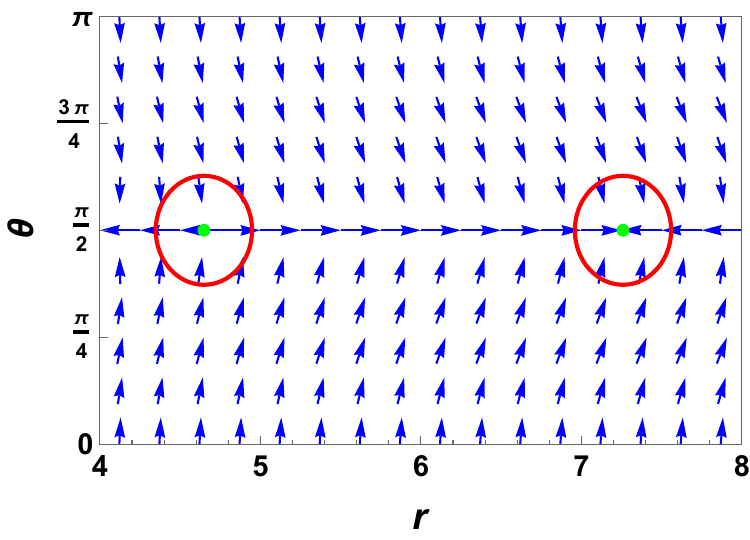}\label{S1t4}}
    \caption{Unit vector $\vec{n}_2=(\phi^r_2,\phi^\theta_2)/||\vec{\phi}_2||$ constructed from $E_2$ on ($r,\,\theta$) plane in the like weak charge regime with $q=1.2$. (a) Angular momentum $L=4.2<L_{ISCO}$. (b) $L=4.5>L_{ISCO}$. Two green zero points locate at $r=4.6481$ and 7.2591. Here $L_{ISCO}=4.3881$.}
    \label{S1top2}
\end{figure}

Apart from the situation of unit vector $\vec{n}_1$, we also need to focus on $\vec{n}_2= \vec{\phi}_2/||\vec{\phi}_2||$ generated from $E_2$. In this case, based on the angular momentum of another ISCO, two dissimilar topological configurations of $\vec{n}_2$ can be depicted in Figs. \ref{S1t3} and \ref{S1t4}, respectively. Analogous to the case of $E_1$ in the regime $0<qQ<M$, such a pattern of vector configurations using the previous conclusions $\club$, $\spad$, and $\loze$ exhibits the vanishing topological number $W_2=0$ before and after $L_{ISCO}$. Combining the result from $E_1$ and $E_2$, the total topological number is $W_{tot}=W_1+W_2=0$ in the like weak charge regime, which is consistent with the result in Table \ref{tab1}.

\begin{figure}[htb]
    \centering
    \subfigure[]{\includegraphics[width=6cm]{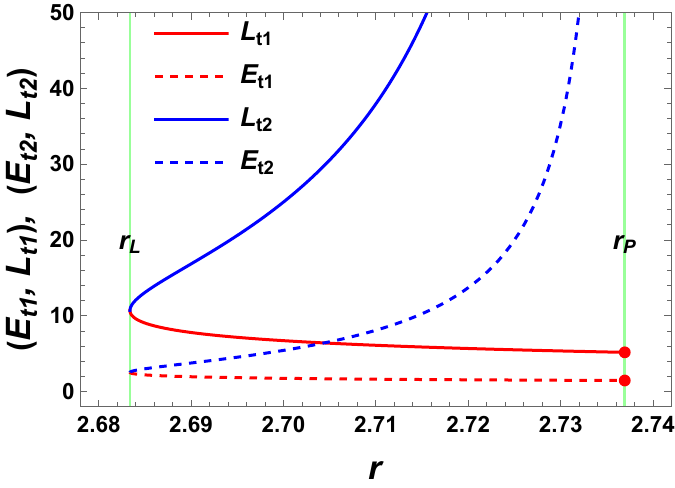}\label{S1ELa}} \hspace{6mm}
    \subfigure[]{\includegraphics[width=6cm]{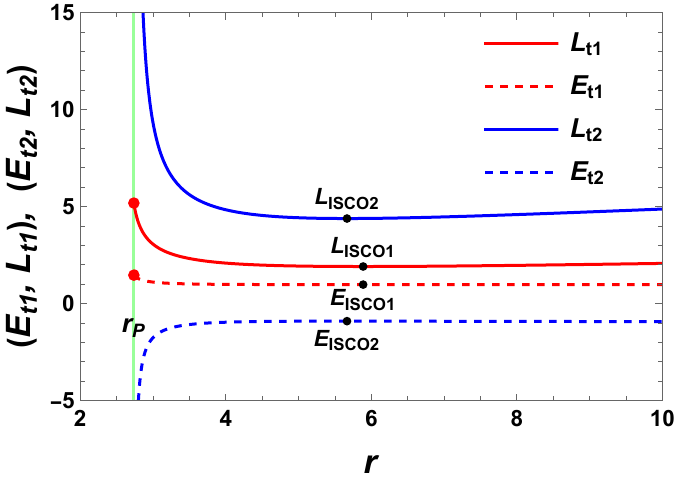}\label{S1ELb}}
    \caption{The solutions $(E_{t1},L_{t1})$ and $(E_{t2},L_{t2})$ of TCOs in the like weak charge regime with $q=1.2$. (a) Interval [$r_L,r_P$]. (b) Interval [$r_P,\infty$). The red point indicates that this curve can be extended and surpass $r_P$. The location of ISCO is denoted as the black point. Here $r_L=2.6833$ and $r_P=2.7369$. }
\end{figure}

The energy and angular momentum of the TCOs, denoted as $(E_{t1},L_{t1})$ and $(E_{t2}, L_{t2})$, are plotted in Figs. \ref{S1ELa} and \ref{S1ELb} corresponding to intervals $[r_L,r_P]$ and $[r_P,\infty)$, respectively. Significantly, it is observed that $E_{t2}$ and $L_{t2}$ are divided into two distinct branches, the one in Fig. \ref{S1ELa} and the other in Fig. \ref{S1ELb}, quite different from $E_{t1}$ and $L_{t1}$, which can be extended from $r_P$ to $r=\infty$. The red points at $r_P$ mean that the corresponding curves are jointed at this point. Besides, we noticed that the solution $(E_{t2},L_{t2})$ can link with $(E_{t1},L_{t1})$ at $r_L$ in Fig. \ref{S1ELa}. Considering the scenario in Fig. \ref{S1ELb}, it is clearly observed that $E_{t1}$, $L_{t1}$, $E_{t2}$, and $L_{t2}$ both exist the minimal value in the range $[r_P, \infty)$ corresponding to two diverse ISCOs. Calculating the constraint relations \eqref{RNISa} and \eqref{RNISb}, the radial radius, energy and angular momentum of these two ISCO can be achieved as
\begin{align}
    &r_{ISCO1}=5.8871, \quad  E_{t1}=0.9835, \quad L_{t1}=1.9160;\\
    &r_{ISCO2}=5.6653, \quad E_{t2}= -0.8991, \quad L_{t2}=4.3881.
\end{align}
It is worth emphasizing that the second ISCO has negative energy $E_{t2}<0$. At the same time, for the solution $(E_{t2},L_{t2})$, we also find that $E_{t2}<0$ within the interval $[r_P,\infty)$.

One important fact is that the combined branch $L_{t1}\cup L_{t2}$ with $r_{ISCO1}$ and the branch $L_{t2}$ with $r_{ISCO2}$ in Figs. \ref{S1ELa} and \ref{S1ELb} can separately interpret the topological number $W_1=0$ and $W_2=0$ of TCO given by Figs. \ref{S1top1} and \ref{S1top2}.

\begin{figure}[hbt]
    \centering
    \includegraphics[width=7cm]{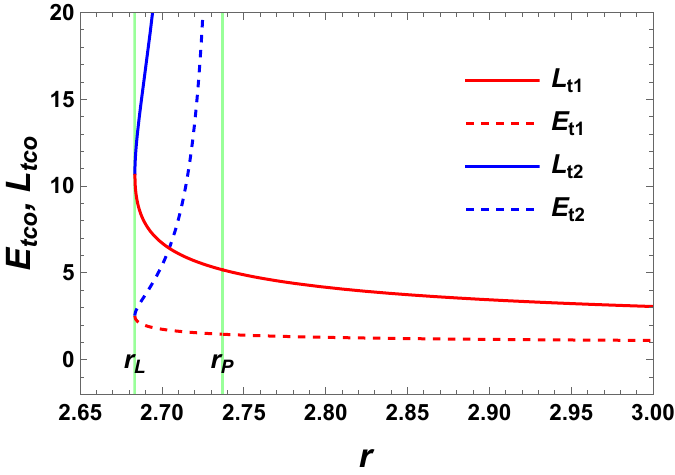}
    \caption{Valid energy $E_{tco}$ and angular momentum $L_{tco}$ of TCO in the like weak charge regime with $q=1.2$. The $E_{t1}$ and $E_{t2}$ together with $L_{t1}$ and $L_{t2}$ are connected at $r_L$. Here $r_L=2.6833$ and $r_P=2.7369$.}
    \label{S1ELc}
\end{figure}

After abandoning the negative energy portion of TCOs and combining two sets of solution, the valid energy $E_{tco}$ and angular momentum $L_{tco}$ are shown as
\begin{align}
    &E_{tco}(r)=\left\{E_{t1}(r_a) \bigcup E_{t2}(r_b) \,\bigg|\, r_a\in [r_L,\infty), r_b\in [r_L,r_P),0<qQ<M  \right\}, \label{Etco3}\\
    &L_{tco}(r)=\left\{L_{t1}(r_a) \bigcup L_{t2}(r_b) \,\bigg|\, r_a\in [r_L,\infty), r_b\in [r_L,r_P),0<qQ<M \right\}, \label{Ltco3}
\end{align}
which are depicted in Fig. \ref{S1ELc}.

Remarkably, the red and blue curves representing $L_{t1}(r)$ and $L_{t2}(r)$ shown in Fig. \ref{S1ELc} exhibit a connection point at $r=r_L$. This unique characteristic is a notable novelty observed within the regime of like weak charge, which distinguishes it from the cases depicted in Figs. \ref{O1ELc} and \ref{O2ELc}, where $E_{t2}$ and $L_{t2}$ do not contribute to $E_{tco}$ and $L_{tco}$. Furthermore, in Fig. \ref{S1ELc}, the curves of $E_{tco}(r)$ and $L_{tco}(r)$ appear as double-valued functions within the range of $[r_L,r_P)$. However, it is essential to emphasize that $E_{tco}$ and $L_{tco}$ are not single, smooth, and continuous functions resulting from the union of \eqref{Etco3} and \eqref{Ltco3}. Instead, they consist of two distinct parts originating from TCOs with $E_{t1}$ and $L_{t1}$ or TCOs with $E_{t2}$ and $L_{t2}$.

\begin{figure}
    \centering
    \subfigure[]{\includegraphics[width=6cm]{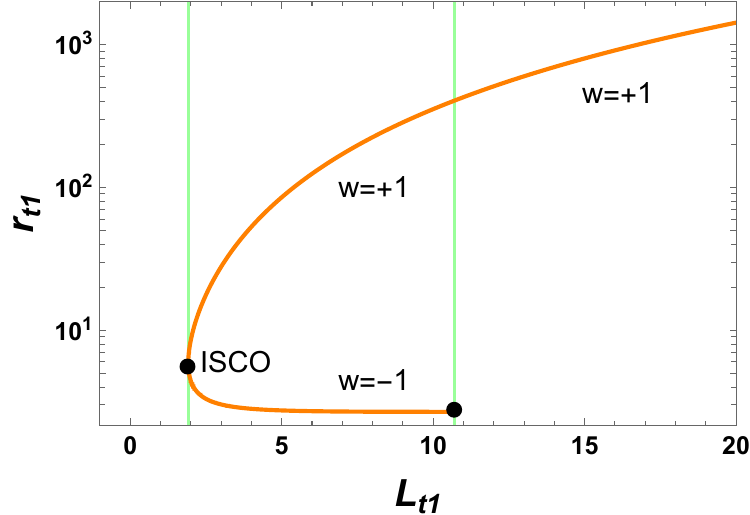}\label{S1r1}} \hspace{6mm}
    \subfigure[]{\includegraphics[width=6cm]{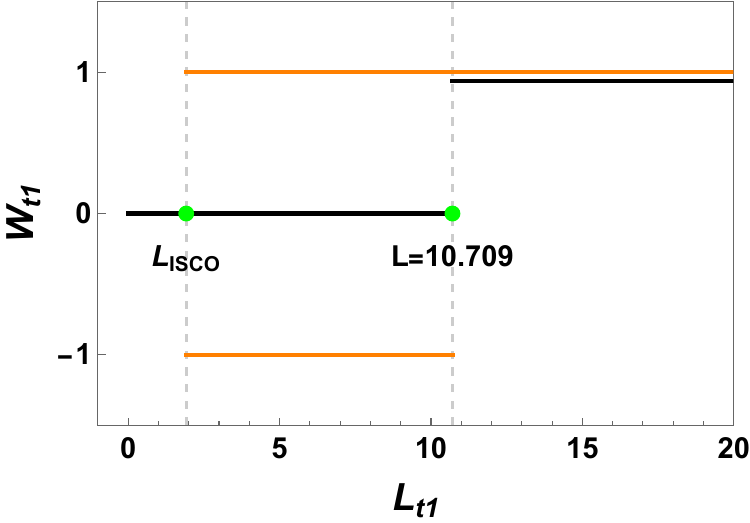}\label{S1w1}}\\
    \subfigure[]{\includegraphics[width=6cm]{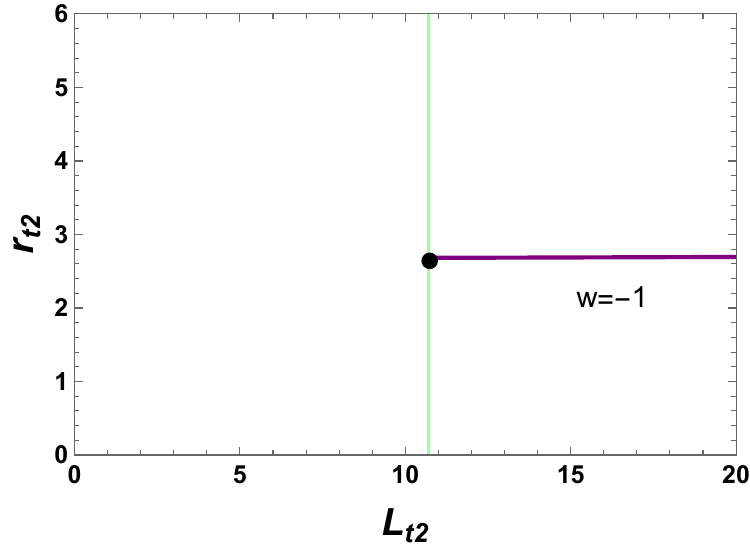}\label{S1r2}} \hspace{6mm}
    \subfigure[]{\includegraphics[width=6cm]{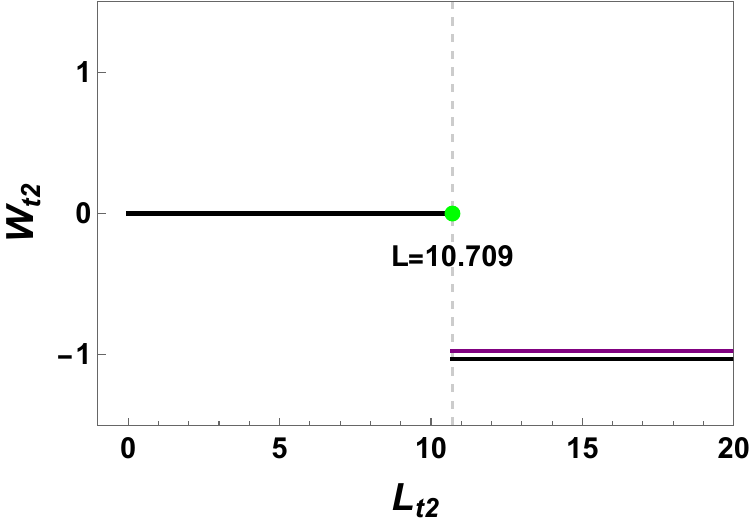}\label{S1w2}}\\
    \subfigure[]{\includegraphics[width=6cm]{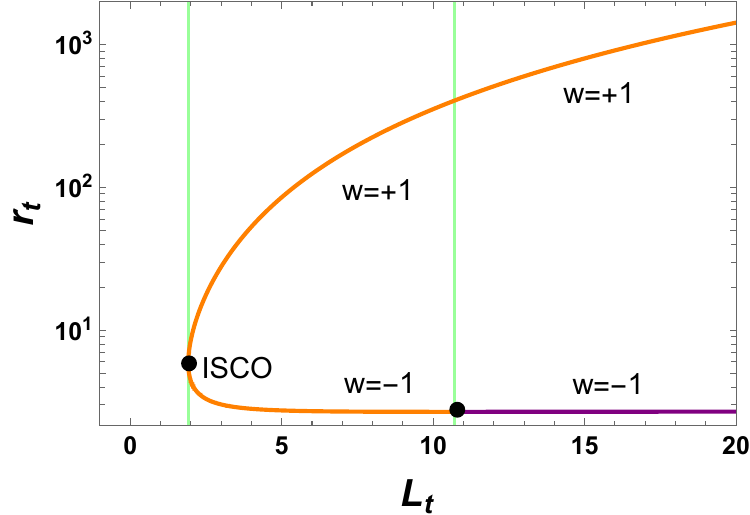}\label{S1rL}} \hspace{6mm}
    \subfigure[]{\includegraphics[width=6cm]{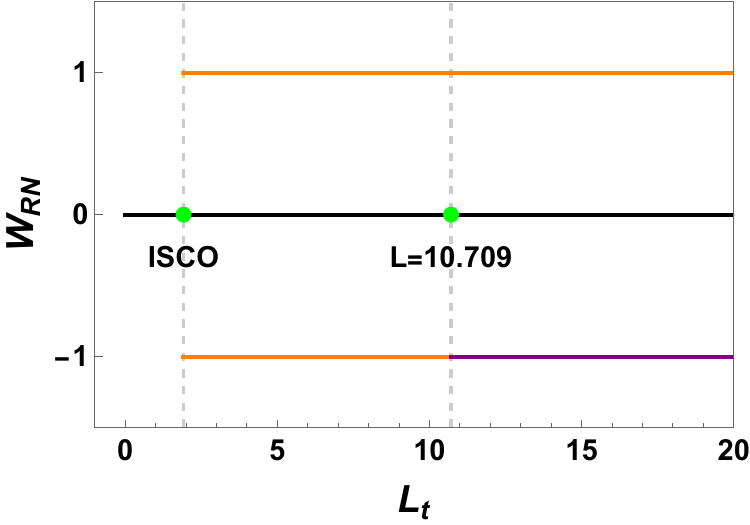}\label{S1wL}}
    \caption{(a) TCO radius $r_{t1}$ as a function of angular momentum $L_{t1}$.  (b) Topological number $W_{t1}$ (black line) as a function of $L_{t1}$. (c) $r_{t2}-L_{t2}$. (d) $W_{t2}-L_{t2}$. (e) $r_t-L_t$. (f) $W_{RN}-L_t$. Here $W=\pm 1$ represent stable and unstable TCOs. The black point denotes ISCO that is a generated bifurcation point. The orange and purple lines stand for TCO branch.}
\end{figure}

Although the topological configurations of the unit vector $\vec{n}_1$ appear quite simple, see Figs. \ref{S1t1} and \ref{S1t2}, the behaviors of $E_{tco}$ and $L_{tco}$ to TCOs suggest that there are additional hidden details within the topological configurations of $\vec{n}_1$ that need to be unveiled. In contrast to the previous cases depicted in Figs. \ref{O1rL} and \ref{O2rL}, where only a single $r_t(L)$ curve was presented, two $r_t(L)$ curves emerge here. We now consider three distinct situations: $r_{t1}(L_{t1})$, $r_{t2}(L_{t2})$, and their combination $r_t(L_t)$. We, respectively, illustrate these situations in Figs. \ref{S1r1}, \ref{S1r2}, and \ref{S1rL}.

Considering the curve $r_{t1}(L_{t1})$ depicted in Fig. \ref{S1r1}, it is evident that the ISCO, denoted by the black point, divides $r_{t1}(L_{t1})$ of the TCOs into two distinct branches. The unstable branch, characterized by radii $r<r_{ISCO}$, originates at $L_{ISCO}$ and terminates at a finite angular momentum value of $L=10.709$. On the other hand, the stable branch, corresponding to larger radii $r>r_{ISCO}$, also starts at $L_{ISCO}$ but extends indefinitely to $L=\infty$. Consequently, the total topological number $W_{t1}$ is equal to zero when $L<L_{ISCO}$ and $L_{ISCO}<L<10.709$. However, when $L>10.709$, $W_{t1}$ becomes +1 due to the presence of a single stable TCO, as indicated by the black line in Fig. \ref{S1w1}. This observation strongly suggests the occurrence of a topological phase transition, where $W_{t1}$ changes from $0$ to $+1$ precisely at $L=10.709$. This phenomenon is remarkably intriguing and has not been previously observed.

Regarding the curve $r_{t2}(L_{t2})$ depicted in Fig. \ref{S1r2}, no ISCO is present, and only one unstable TCO branch appears within the range of $L\in [10.709,\infty)$. Consequently, when $L<10.709$, no TCOs exist, resulting in a vanishing topological number $W_{t2}=0$. However, when $L>10.709$, the topological number becomes $W_{t2}=-1$, as clearly illustrated in Fig. \ref{S1w2}. The transition of the topological number from $W_{t2}=0$ to -1 signifies the occurrence of a topological phase transition precisely at $L=10.709$. It is noteworthy that the topological phase transitions in $W_{t1}(L_{t1})$ and $W_{t1}(L_{t2})$ both occur at the same angular momentum value $L=10.709$.

Next, we examine the combination of $r_{t1}(L_{t1})$ and $r_{t2}(L_{t2})$ in $r_t(L_t)$ as illustrated in Fig. \ref{S1rL}. In this case, when $L<L_{ISCO}$, no TCOs are present. However, as the angular momentum increases within the range $L_{ISCO}<L<10.709$, $r_{t1}(L_{t1})$ exhibits a pair TCO branches. Subsequently, in the interval $L\in[10.709,\infty)$, $r_{t1}(L_{t1})$ and $r_{t2}(L_{t2})$ contribute a stable and an unstable TCO, respectively. As a result, the total topological number $W_{RN}=0$ remains unchanged both before and after $L=10.709$, as depicted in Fig. \ref{S1wL}. This result is in agreement with $W_{tot}=0$ in the Table \ref{tab1} which do not take into account the negative energy.

\subsection{Like Strong Charge Regime: $0<M<qQ$}

In this regime of like strong charge case $0<M<qQ$, the Coulomb repulsive force may surpass the attraction of the RN black hole. Additionally, as indicated in Table \ref{tab1}, the total topological number is $W_{tot}=-1$ within this specific regime. Notably, this differs from the weak charge regimes of $0<|qQ|<M$ that we have previously discussed.

It is noteworthy that we further subdivide this regime into two distinct subcases: (1) $M<qQ<q_t Q$ and (2) $q_t Q<qQ<\infty$, based on the differing energy and angular momentum of the TCOs depicted in Figs. \ref{figs2} and \ref{figs3}. Here, $q_t$ represents a special charge-to-mass ratio, and its analytical expression is given by \cite{Pugliese2011py}
\begin{equation}
    q_t= \frac{1}{\sqrt{2}Q} \sqrt{5M^2-4Q^2+\sqrt{25 M^2-24Q^2}}.
\end{equation}

\subsubsection{Subcase: $M<qQ<q_tQ$}

\begin{figure}[htb]
    \centering
    \includegraphics[width=7cm]{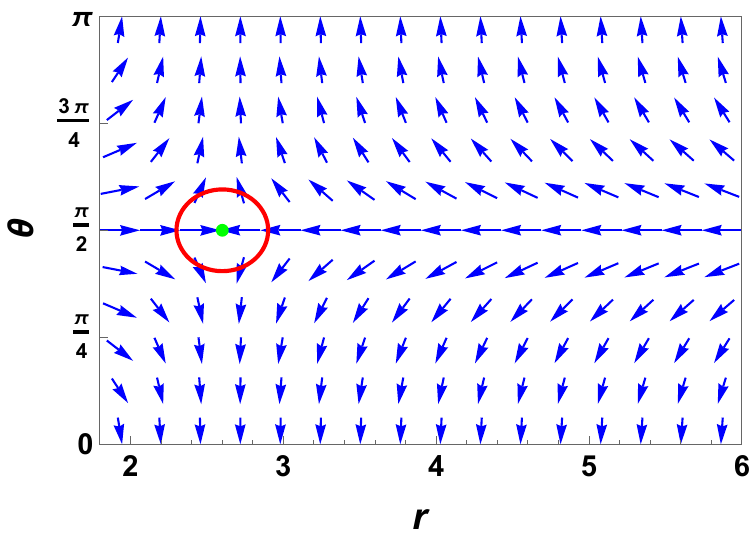}
    \caption{Unit vector $\vec{n}_1=(\phi^r_1,\phi^\theta_1)/||\vec{\phi}_1||$ constructed from $E_1$ with angular momentum $L=10$ on $(r,\theta)$ plane in the first subcase $M<qQ<q_tQ$ with $q=2.1$. The topological configuration is similar for arbitrary $L>0$. One green zero point represents an unstable TCO and locates at $r=2.6009$.}
    \label{S2t1}
\end{figure}

We now investigate the first subcase, where $M<qQ<q_tQ$, and consider a specific charge-to-mass ratio $q=2.1$. In Fig. \ref{S2t1}, we present the topological configuration of the unit vector $\vec{n}_1$ for $E_1$ with $L=10$ in the $(r,\theta)$ plane. Notably, one can see a zero point of $\vec{n}_1$, which is indicated by the green marker and surrounded by a closed red path.

It holds true that within this subcase of the like strong charge regime, only one type of topological configuration of the unit $\vec{n}_1$ can be found, regardless of the specific value of the angular momentum. In this configuration, the vector arrows flow towards the zero point along the $r$-direction and outwards from it along the $\theta$-direction. From the conclusion $\loze$, we can conclude that the zero point possesses a topological charge $W_1=-1$ for $E_1$, corresponding to an unstable TCO. Obviously, this configuration is exactly consistent with our asymptotic analysis of the $\vec{\phi}_1$ constructed from $E_1$ at $\partial \Sigma$ depicted in Fig. \ref{Bca}.

\begin{figure}[hbt]
    \centering
    \subfigure[]{\includegraphics[width=6cm]{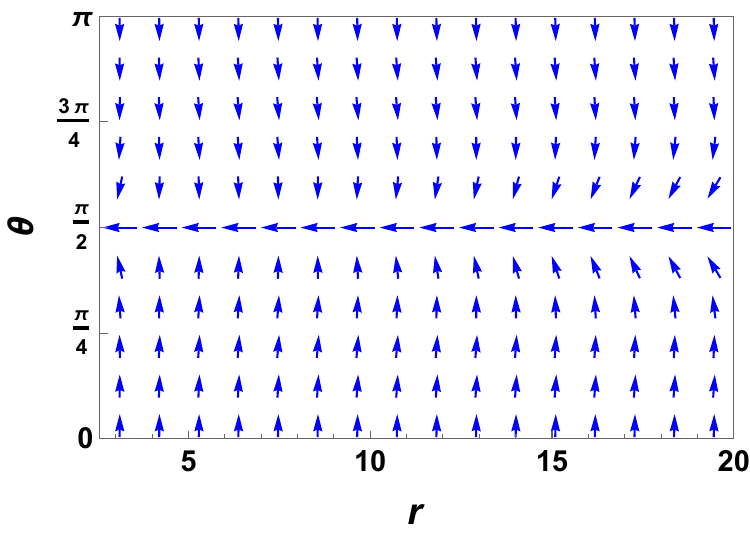}\label{S2t2}}\hspace{6mm}
    \subfigure[]{\includegraphics[width=6cm]{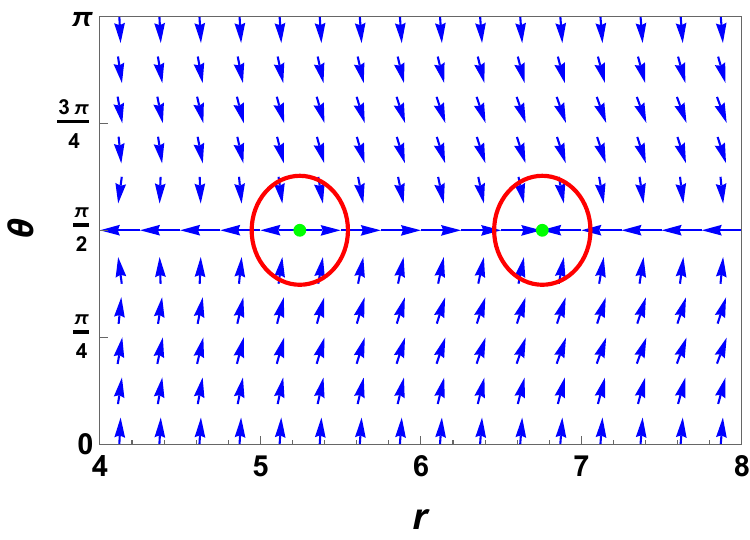}\label{S2t3}}
    \caption{Unit vector $\vec{n}_2=(\phi^r_2,\phi^\theta_2)/||\vec{\phi}_2||$ constructed from $E_2$ on ($r,\,\theta$) plane in the first subcase $M<qQ<q_tQ$ with $q=2.1$. (a) Angular momentum $L=5.15<L_{ISCO}$. (b) $L=5.2>L_{ISCO}$. Two green zero points locate at $r=5.2455$ and 6.7564. Here $L_{ISCO}=5.1587$.}
    \label{S2top2}
\end{figure}

We further examine the scenario of vector $\vec{n}_2$ that is calculated from $E_2$ and the corresponding topological configuration of unit vector $\vec{n}_2$ can be illustrated in Figs. \ref{S2t2} and \ref{S2t3}. As we have seen earlier in other charge regimes, these two configurations are split by angular momentum of ISCO. In the former case with $L=5.15<L_{ISCO}$, no any zero points exist and this results in the topological number $W=0$. While for the latter case with $L>L_{ISCO}$, we can observe that two green zero points emerge concurrently, and they separately possess topological number $W=-1$ and +1 according to previous conclusions $\loze$ and $\spad$. Summing over these two number, the topological number for the second configuration of unit vector also remain $W=0$. As a result, $W_2=0$ is expected for the vector $\vec{n}_2$ built from $E_2$. Combining the outcome from $E_1$ and $E_2$, we end up with total topological number $W_{tot}=W_1+W_2=-1$, which keeps same with the result in Table \ref{tab1}.

\begin{figure}[htb]
    \centering
    \subfigure[]{\includegraphics[width=6cm]{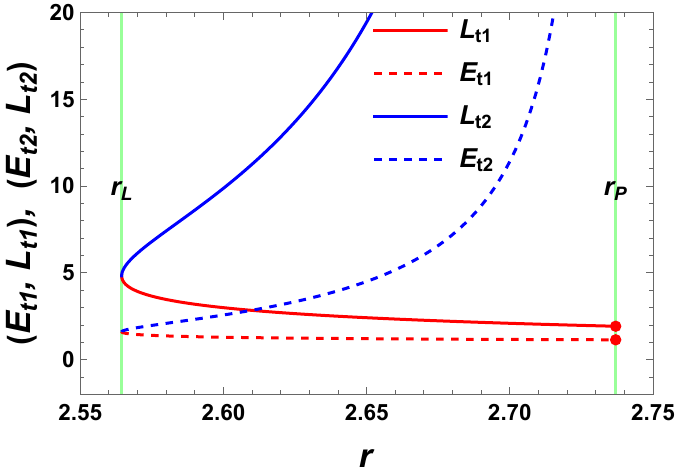}\label{S2ELa}} \hspace{6mm}
    \subfigure[]{\includegraphics[width=6cm]{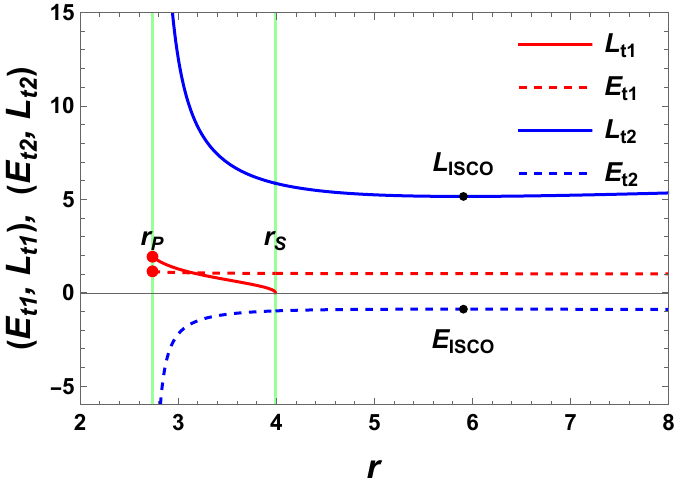}\label{S2ELb}}
    \caption{The solutions $(E_{t1},L_{t1})$ and $(E_{t2},L_{t2})$ of TCOs in the first subcase of the like strong charge regime with $q=2.1$. (a) Interval [$r_L,r_P$]. (b) Interval [$r_P,\infty$). The red point indicates that this curve can be extended and surpass $r_P$. The location of ISCO is denoted as the black point. Here $r_L=2.5644$, $r_S=3.9898$, and $r_P=2.7369$. }
    \label{figs2}
\end{figure}

Let us now turn our attention to the energy and angular momentum of the TCOs. We plot $(E_{t1},L_{t1})$ and $(E_{t2},L_{t2})$ in Figs. \ref{S2ELa} and \ref{S2ELb} corresponding to intervals $[r_L,r_P]$ and $[r_P,\infty)$, respectively. In Fig. \ref{S2ELa}, We see that $L_{t1}$ and $E_{t1}$ both are positive and initiate at $r_L$, whereas they have dissimilar ending points. Specifically, $L_{t1}$ and $E_{t1}$ extends from $r_P$ to $r_s$ and $\infty$, respectively. Note that $L_{t1}=0$ at the position $r_S$. Furthermore, the TCOs characterized by positive $E_{t1}$ and vanishing $L_{t1}$ at $r_s$ do not exhibit any relative rotation with respect to distant observers, which is exactly the static point orbit observed in Ref. \cite{Wei2023bgp} for a uncharged particle. Here the red points on the red solid and dashed curves indicate that the curves can continue to stretch. As for $E_{t2}$ and $L_{t2}$, they are divided into two sectors against radius $r_P$ displayed in Figs. \ref{S2ELa} and \ref{S2ELb}, respectively. In the left figure, both $L_{t2}$ and $E_{t2}$ are positive and they link with $L_{t1}$ and $E_{t1}$ at the starting point $r_L$. These two curves increase with the TCO radius $r$. However, in the right figure, $E_{t2}$ becomes negative and they separately show an extreme point corresponding to ISCO. The radial radius, energy, and angular momentum of ISCO can be gained from the Eq. \eqref{RNISb}
\begin{equation}
    r_{ISCO}=5.9067, \quad E_{t2}=-0.9561, \quad L_{t2}=2.7728.
\end{equation}
Notice that the energy $E_{t2}$ of this ISCO  is negative.

One important fact is that the combined branch $L_{t1}\cup L_{t2}$ and the branch $L_{t2}$ with $r_{ISCO}$ in Figs. \ref{S2ELa} and \ref{S2ELb} can separately interpret the topological number $W_1=-1$ and $W_2=0$ of TCO given by Figs. \ref{S2t1} and \ref{S2top2}.

\begin{figure}[htb]
    \centering
    \includegraphics[width=7cm]{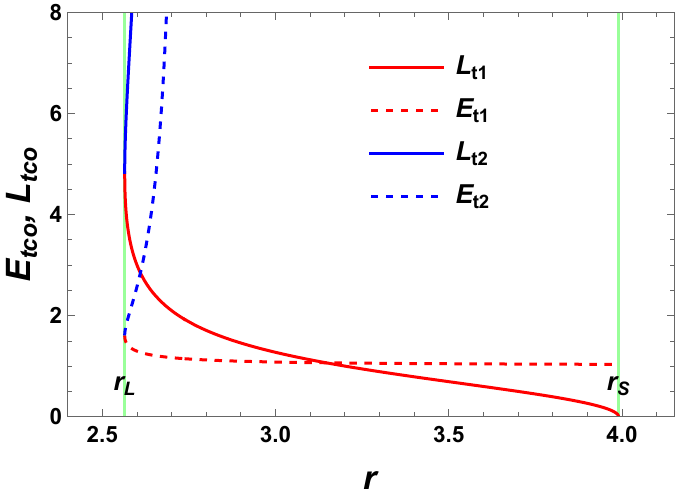}
    \caption{Valid energy $E_{tco}$ and angular momentum $L_{tco}$ of TCO in the first subcase of the like strong charge regime with $q=2.1$. The $E_{t1}$ and $E_{t2}$ together with $L_{t1}$ and $L_{t2}$ are connected at $r_L$ and terminate at $r_S$. Here $r_P=2.7369$ and $r_S=3.9898$.}
    \label{S2ELc}
\end{figure}

After excluding the negative energy portion, the physically meaningful energy $E_{tco}$ and angular momentum $L_{tco}$ of the TCOs can be defined as follows:
\begin{align}
    &E_{tco}(r)=\left\{E_{t1}(r_a) \bigcup E_{t2}(r_b) \,\bigg|\, r_a\in [r_L,r_S], r_b\in [r_L,r_P),M<qQ<q_t Q  \right\},  \\
    &L_{tco}(r)=\left\{L_{t1}(r_a) \bigcup L_{t2}(r_b) \,\bigg|\, r_a\in [r_L,r_S], r_b\in [r_L,r_P),M<qQ<q_t Q \right\}.
\end{align}
The $E_{tco}$ and $L_{tco}$ are illustrated in Fig. \ref{S2ELc}. In this figure, $E_{t1}$ and $E_{t2}$, as well as $L_{t1}$ and $L_{t2}$, are connected at $r=r_L$. When comparing with them depicted in Fig. \ref{S1ELc}, we observe that in the present case, $E_{tco}$ and $L_{tco}$ must terminate at $r_S$ due to the vanishing angular momentum rather than extend to radial infinity at $r=\infty$.

\begin{figure}[hbt]
    \centering
    \subfigure[]{\includegraphics[width=6cm]{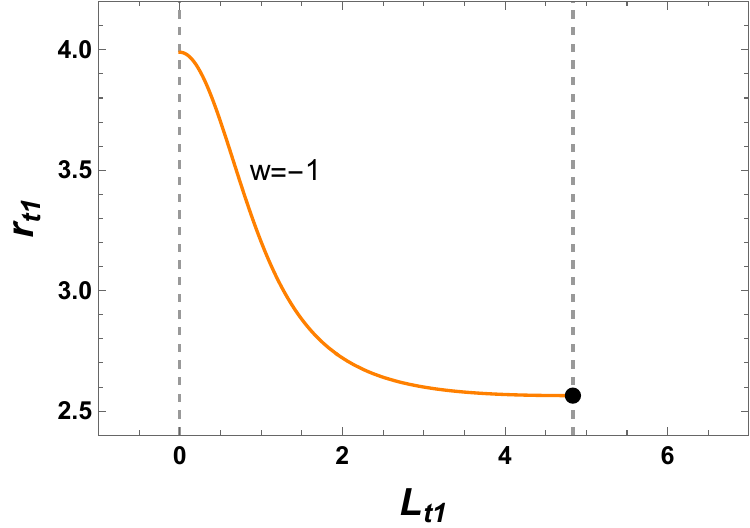}\label{S2r1}} \hspace{6mm}
    \subfigure[]{\includegraphics[width=6cm]{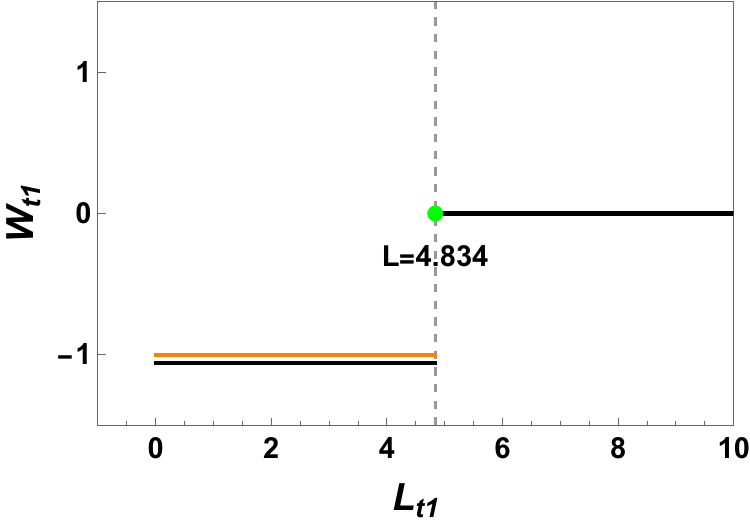}\label{S2w1}}\\
    \subfigure[]{\includegraphics[width=6cm]{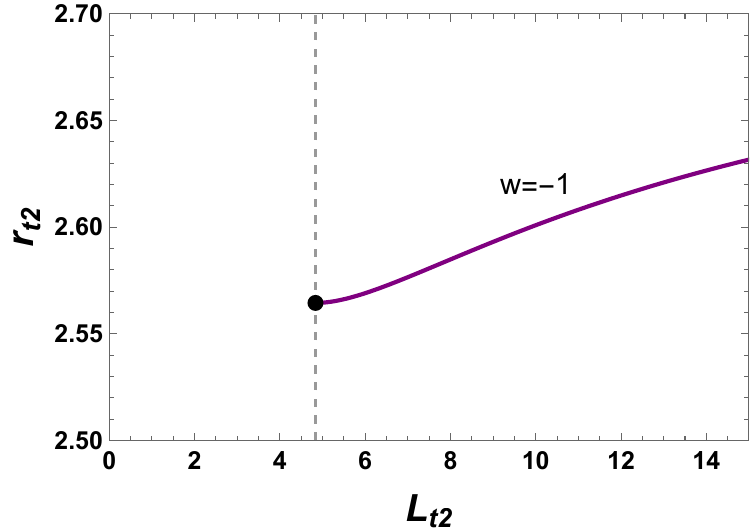}\label{S2r2}} \hspace{6mm}
    \subfigure[]{\includegraphics[width=6cm]{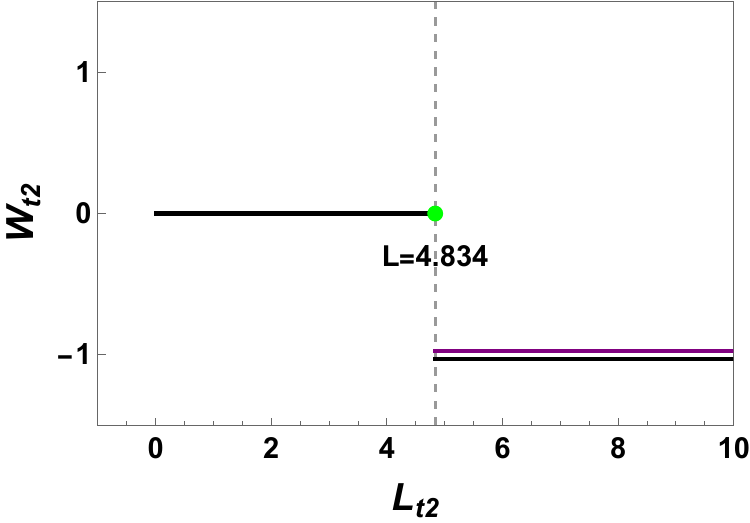}\label{S2w2}}\\
    \subfigure[]{\includegraphics[width=6cm]{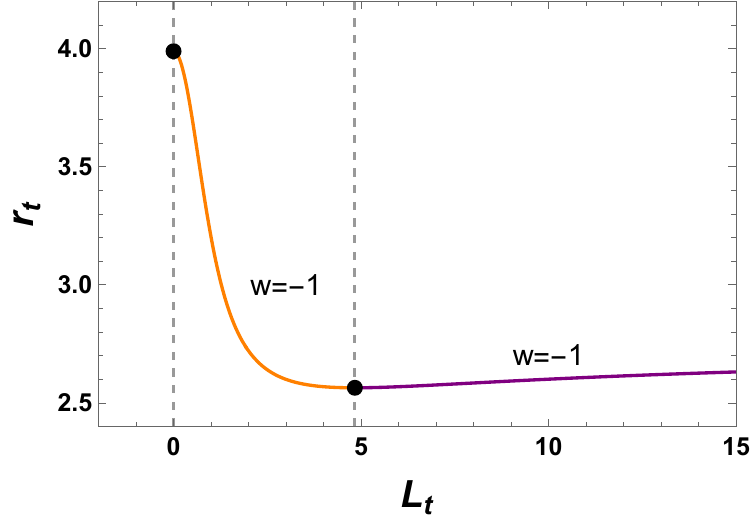}\label{S2rL}} \hspace{6mm}
    \subfigure[]{\includegraphics[width=6cm]{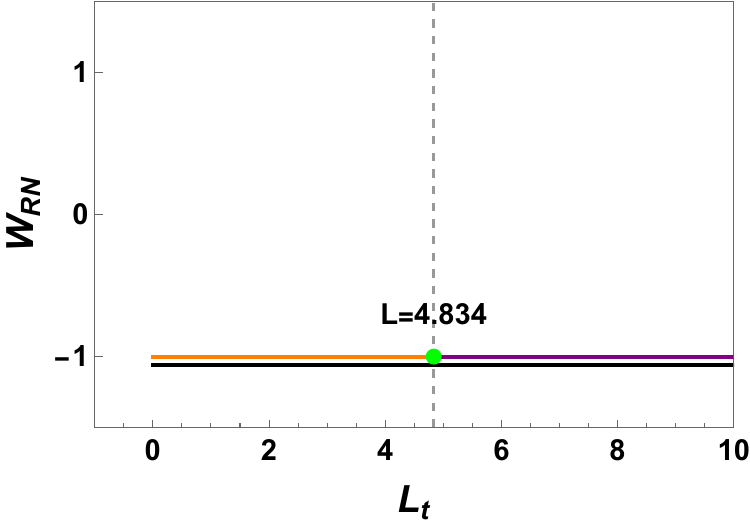}\label{S2wL}}
    \caption{(a) TCO radius $r_{t1}$ as a function of angular momentum $L_{t1}$.  (b) Topological number $W_{t1}$ (black line) as a function of $L_{t1}$. (c) $r_{t2}-L_{t2}$. (d) $W_{t2}-L_{t2}$. (e) $r_t-L_t$. (f) $W_{RN}-L_t$. Here $W= -1$ represents unstable TCOs. The orange and purple lines stand for TCO branch.}
\end{figure}

Here, we aim to examine the relationship between the TCO radius and the angular momentum based on $E_{tco}$ and $L_{tco}$. As $L_{tco}$ is an union of both $L_{t1}$ and $L_{t2}$, we need to consider three scenarios: $r_{t1}(L_{t1})$, $r_{t2}(L_{t2})$, and their union $r_t(L_t)$, respectively. In Fig. \ref{S2r1}, $r_{t1}(L_{t1})$ initiates at $L=0$ and terminates at $L=4.834$. It represents an unstable TCO branch that exhibits a topological number $W_{t1}=-1$ for $L<4.834$. However, in the absence of TCOs for $L>4.834$, the topological number becomes $W_{t1}=0$, as depicted in Fig. \ref{S2w1}. Consequently, a topological phase transition occurs at the angular momentum value of $L=4.834$, where the topological number turns from $W_{t1}=-1$ to $0$.

In the previous case, we also have the topological phase transition as depicted in Fig. \ref{S1r1}. However, it should be noted that the existence of the ISCO differs from the current case. In contrast, the $r_{t2}(L_{t2})$ curve in Fig. \ref{S2r2} originates at a non-zero angular momentum value of $L=4.834$ and extends indefinitely to $L=\infty$. This TCO branch is also unstable, and the corresponding topological number $W_{t2}=0$ and $-1$ before and after $L=4.834$ as illustrated in Fig. \ref{S2w2}. The occurrence of a topological phase transition in the $W_{t2}$-$L_{t2}$ diagram leads to a variation in the topological number from $W_{t2}=0$ to $-1$. Remarkably, this phenomenon exhibits a striking analogy to the situation observed in $W_{t1}$ while with a different sign.

The final result for the $r_t$-$L_t$ relation is simply the combination of $r_{t1}(L_{t1})$ and $r_{t2}(L_{t2})$, as shown in Fig. \ref{S2rL}. It is important to note that both of these TCO branches represent the unstable TCOs. Consequently, the total topological number remains constant $W_{RN}=-1$, as indicated by the black line in Fig. \ref{S2wL}. This result is consistent with the result $W_{tot}=-1$ obtained from Table \ref{tab1} where we do not take into account the negative energy.

\subsubsection{Subcase: $q_tQ<qQ<\infty$}

In the second subcase $q_t Q<qQ<\infty$, we consider a specific charge-to-mass ratio example of $q=3.8$ to investigate the topological configuration of TCOs. Referring to the results in Table \ref{tab1}, we have already established that the total topological number remains constant $W_{tot}=-1$ in the like strong charge regime. Therefore, this subcase also exhibits an identical topological number.

\begin{figure}[htb]
    \centering
    \includegraphics[width=7cm]{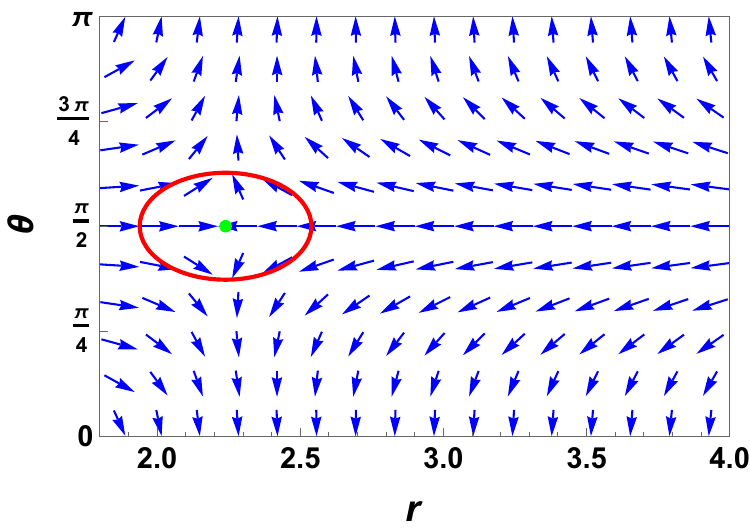}
    \caption{Unit vector $\vec{n}_1=(\phi^r_1,\phi^\theta_1)/||\vec{\phi}_1||$ constructed from $E_1$ with angular momentum $L=2.9$ on $(r,\,\theta)$ plane in the second subcase $q_tQ<qQ<\infty$ with $q=3.8$. The topological configuration is similar for arbitrary positive angular momentum. The green zero point represents an unstable TCO and locates at $r=2.2389$.}
    \label{S3t1}
\end{figure}

In Fig. \ref{S3t1}, we present the unit vector $\vec{n}_1$ for $E_1$ in the $(r, \theta)$ plane for $L=2.9$. A zero point marked in green at $r=2.2389$ is observed. Through further adjustments of $L$, we find that only a single zero point exists in the $(r, \theta)$ plane. This observation suggests that the topological configuration with a single TCO is the unique possibility for any positive angular momentum. The vector arrows in the vicinity of the zero point flow into it along the $r$-direction and out of it along the $\theta$-direction. Based on the conclusion $\loze$, it is evident that this zero point represents an unstable TCO, and its topological number is $W_1=-1$.

\begin{figure}[htb]
    \centering
    \subfigure[]{\includegraphics[width=6cm]{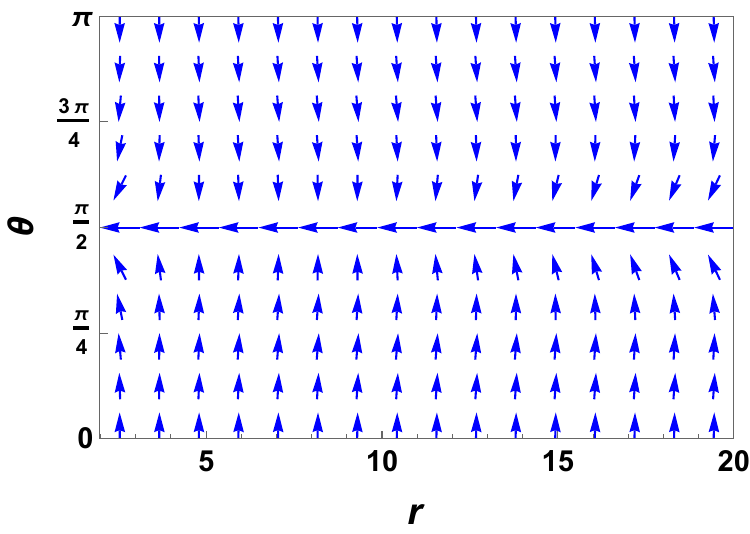}\label{S3t2}} \hspace{6mm}
    \subfigure[]{\includegraphics[width=6cm]{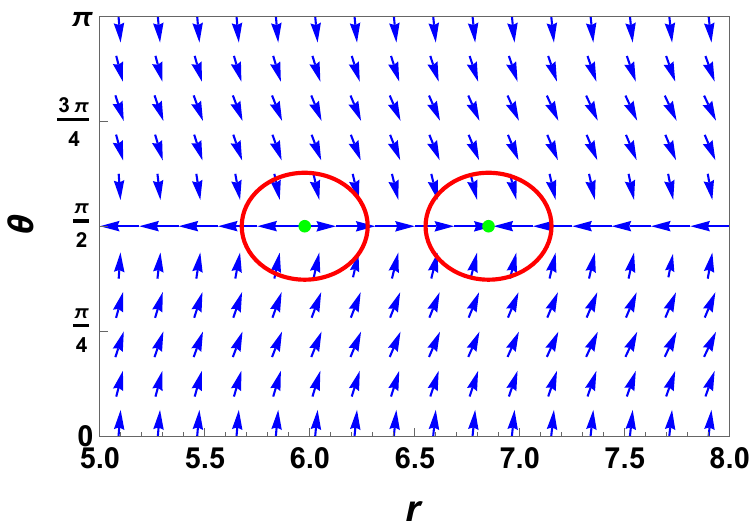}\label{S3t3}}
    \caption{Unit vector $\vec{n}_2=(\phi^r_2,\phi^\theta_2)/||\vec{\phi}_2||$ constructed from $E_2$ on ($r,\,\theta$) plane in the second subcase $q_tQ<qQ<\infty$  with $q=3.8$. (a) Angular momentum $L=6.52<L_{ISCO}$. (b) $L=6.55>L_{ISCO}$. Two green zero points locate at $r=5.9755$ and 6.8511. Here $L_{ISCO}=6.5356$.}
    \label{S3top2}
\end{figure}

Let us move to the situation of the unit vector $\vec{n}_2$ built from $E_2$. The corresponding topological configurations are depicted in Figs. \ref{S3t2} and \ref{S3t3}. In Fig. \ref{S3t2}, we can see that no zero point can be found in $(r,\theta)$ plane when $L=6.52<L_{ISCO}$, and the vector arrows always point to the left at $\theta=\pi/2$. It implies that the topological number in this case is $W_2=0$. For the second configuration with $L=6.55>L_{ISCO}$, unit vector $\vec{n}_2$ exists two zero points marked in green and their radial coordinates are $r=5.9755$ and 6.8511, respectively. In addition, applying the conclusions $\loze$  and $\spad$ earlier, these two zero points separately have topological number $W=-1$ and +1. Thereby, the eventually topological number remain the constant $W_2=0$ for the second configuration. As a result, $W_{tot}=W_1+W_2=-1$ in the regime of $q_tQ<M<\infty$, which coincide with the result given by Table \ref{tab1}.

\begin{figure}[htb]
    \centering
    \subfigure[]{\includegraphics[width=6cm]{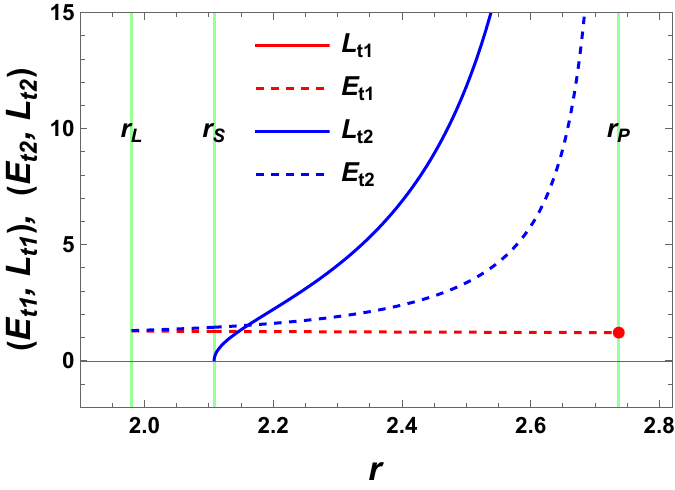}\label{S3ELa}} \hspace{6mm}
    \subfigure[]{\includegraphics[width=6cm]{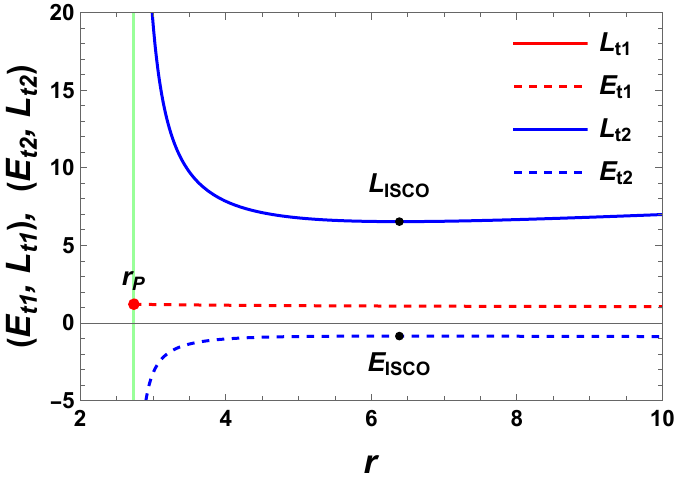}\label{S3ELb}}
    \caption{The solutions $(E_{t1},L_{t1})$ and $(E_{t2},L_{t2})$ of TCOs in the second subcase of the like strong charge regime with $q=3.8$. (a) Interval [$r_L,r_P$]. (b) Interval [$r_P,\infty$). The $L_{t1}$ is an imaginary number and disappears in the figure. The red point indicates that this curve can be extended and surpass $r_P$. The location of ISCO is denoted as the black point. Here $r_L=1.98$, $r_S=2.1087$, and $r_P=2.7369$.}
    \label{figs3}
\end{figure}

We now turn our attention to the energy and angular momentum of TCOs, namely $(E_{t1},L_{t1})$ and $(E_{t2},L_{t2})$, as depicted in Figs. \ref{S3ELa} and \ref{S3ELb}. These two figures correspond to intervals $[r_L,r_P]$ and $[r_P,\infty)$, respectively. A particularly unusual situation arises where $L_{t1}$ becomes an imaginary number and is absent in Fig. \ref{S3ELa} and \ref{S3ELb}, while $E_{t1}$ remains a positive real number that starts at $r_L$ and ends at $r=\infty$. The presence of an imaginary $L_{t1}$ indicates that TCOs characterized by $L_{t1}$ and $E_{t1}$ do not contribute to the topological configuration of the vector. On the other hand, positive $L_{t2}$ can be divided into two distinct sectors. One part of $L_{t2}$ monotonically increases from $r_S$ to $r_P$ shown in Fig. \ref{S3ELa}, while the other part has a minimal value and extends from $r_P$ to radial infinity $r=\infty$, as shown in Fig. \ref{S3ELb}. The energy $E_{t2}$ exhibits increasing behavior both in the intervals $[r_L,r_P)$ and $(r_P,\infty)$, and takes positive and negative values in the former and latter intervals, respectively. The radial radius, energy, and angular momentum of ISCO displayed in Fig. \ref{S3ELb} can be achieved from the constraint relation \eqref{RNISb}
\begin{equation}
    r_{ISCO}=6.3849, \quad E_{t2}=-0.9574, \quad L_{t2}=2.7954.
\end{equation}
Note that this particular ISCO possesses a negative energy $E_{t2}<0$.

One important fact is that the monotonously branch $L_{t2}$ in Fig. \ref{S3ELa} and the branch $L_{t2}$ with $r_{ISCO}$ in Fig. \ref{S2ELb} can separately interpret the topological number $W_1=-1$ and $W_2=0$ of TCO given by Figs. \ref{S3t1} and \ref{S3top2}.

\begin{figure}[htb]
    \centering
    \includegraphics[width=7cm]{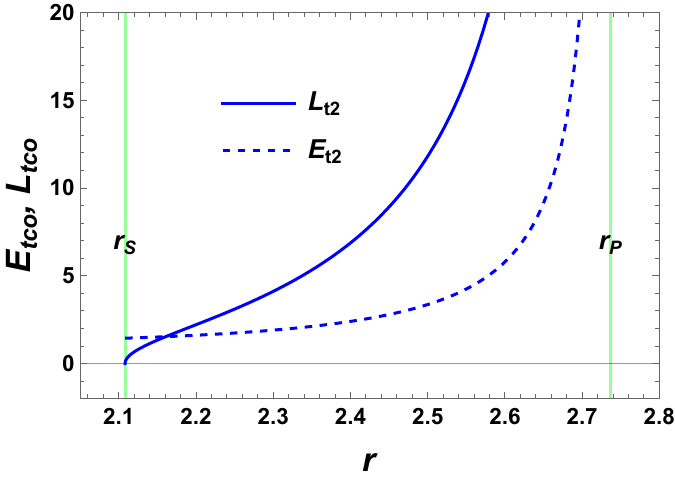}
    \caption{Valid energy $E_{tco}$ and angular momentum $L_{tco}$ of TCOs in the second subcase of the like strong charge regime with $q=3.8$. Here $r_S=2.1087$ and $r_P=2.7369$.}
    \label{S3ELc}
\end{figure}

After excluding the imaginary $L_{t1}$ and the negative energy portion of $E_{t2}$, the valid energy $E_{tco}$ and angular momentum $L_{tco}$ can be defined as follows
\begin{align}
    &E_{tco}(r)=\left\{E_{t1}(r_a) \bigcup E_{t2}(r_b)  \,\bigg|\, r_a\in \varnothing,\, r_b\in [r_S,r_P),\, q_tQ<qQ<\infty \right\}, \label{valE3} \\
    &L_{tco}(r)=\left\{L_{t1}(r_a) \bigcup L_{t2}(r_b) \,\bigg|\, r_a\in \varnothing,\, r_b\in [r_S,r_P),\, q_tQ<qQ<\infty \right\}. \label{valL3}
\end{align}
They are depicted in Fig. \ref{S3ELc} and exhibit a monotonic increase from $r_S$ to $r_P$. Additionally,  $L_{t2}$ equals to zero at the radius $r_S$.

\begin{figure}[htb]
    \centering
    \subfigure[]{\includegraphics[width=6cm]{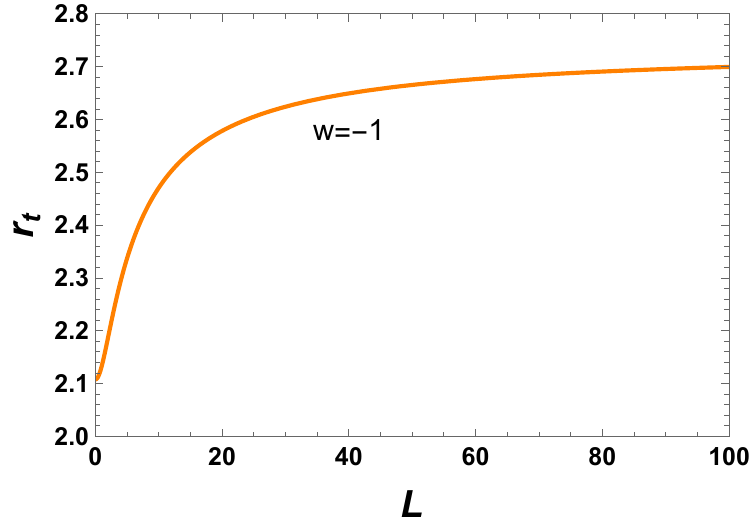}\label{S3rL}} \hspace{6mm}
    \subfigure[]{\includegraphics[width=6cm]{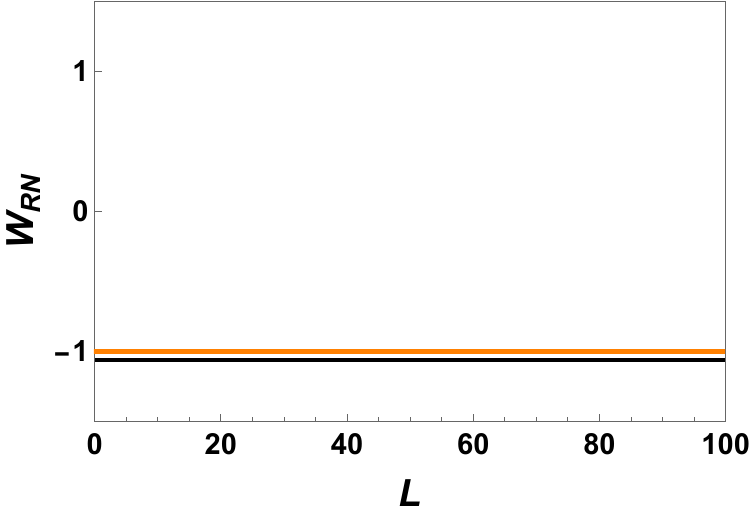}\label{S3wL}}
    \caption{(a) TCO radius $r_t$ as a function of angular momentum $L$.  (b) Topological number $W_{RN}$ as a function of $L$. Here $W=-1$ represents unstable TCO. The orange line stands for TCO branch.}
\end{figure}

Now, we turn our attention to the relationship between the TCO radius $r_t$ and the angular momentum $L$ according to $E_{tco}$ and $L_{tco}$, and then investigate the topological number $W_{RN}$ as a function of $L$. The representation of $r_t$-$L$ in Fig. \ref{S3rL} consistently increases from $L=0$ to $L=\infty$, indicating the absence of any ISCO in this subcase. Moreover, this TCO branch corresponds to an unstable TCO. Consequently, the total topological charge remains constant, i.e., $W_{RN}=-1$, in all the parameter space. While in the previous subcase $M<qQ<q_tQ$, a topological phase transition can occur in $W_{t1}(L_{t1})$ and $W_{t2}(L_{t2})$ shown in Figs. \ref{S2w1} and \ref{S2w2}, we do not observe such a transition in the $q_tQ<qQ<\infty$ subcase. The result of $W_{RN}=-1$ is consistent with $W_{tot}=-1$ presented from Table \ref{tab1} where we do not take into account the negative energy.

\begin{figure}[H]
    \vspace{4mm}
    \centering
    \subfigure[]{\includegraphics[width=6cm]{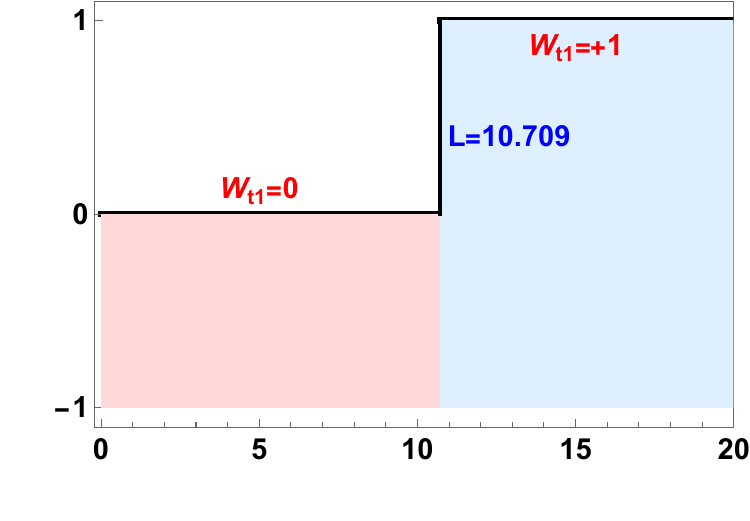}\label{TPT1a}}\hspace{6mm}
    \subfigure[]{\includegraphics[width=6cm]{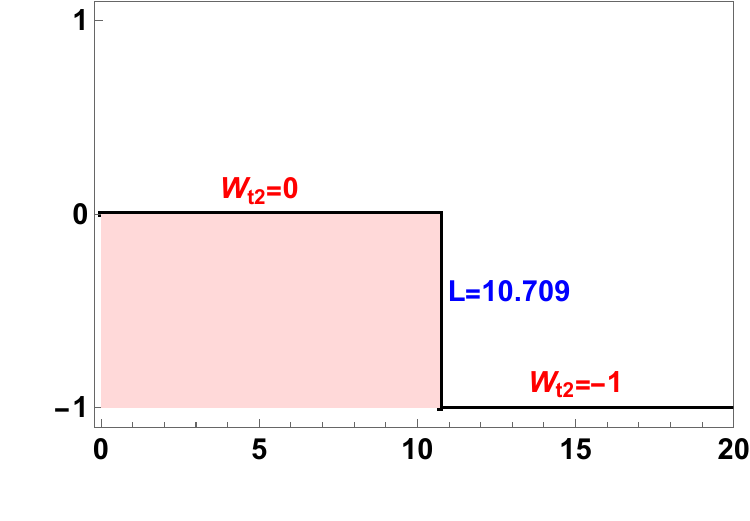}\label{TPT1b}}
    \caption{Topological phase transition of TCO within the charge regime $0<qQ<M$. The topological numbers of the TCO branches $r_{t1}$ and $r_{t2}$ change from $W_{t1}=0$ to +1 and from $W_{t2}=0$ to -1 at $L=10.709$, respectively.} \label{TPT1}
\end{figure}

\begin{figure}[H]
    \centering
    \subfigure[]{\includegraphics[width=6cm]{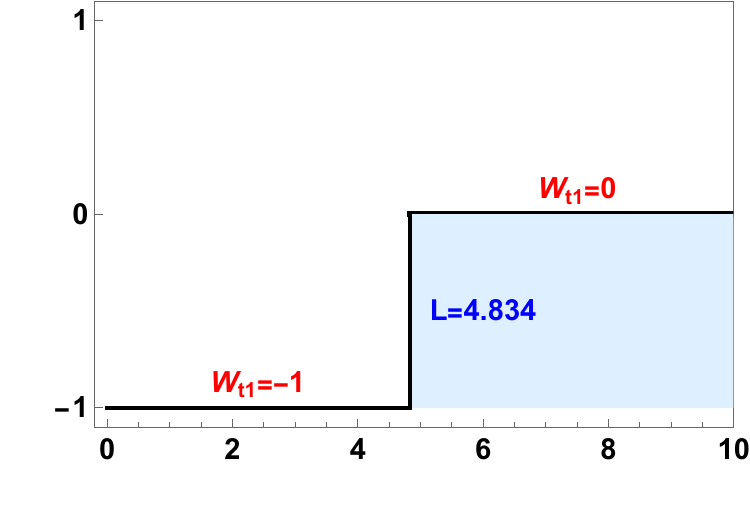}\label{TPT2a}}\hspace{6mm}
    \subfigure[]{\includegraphics[width=6cm]{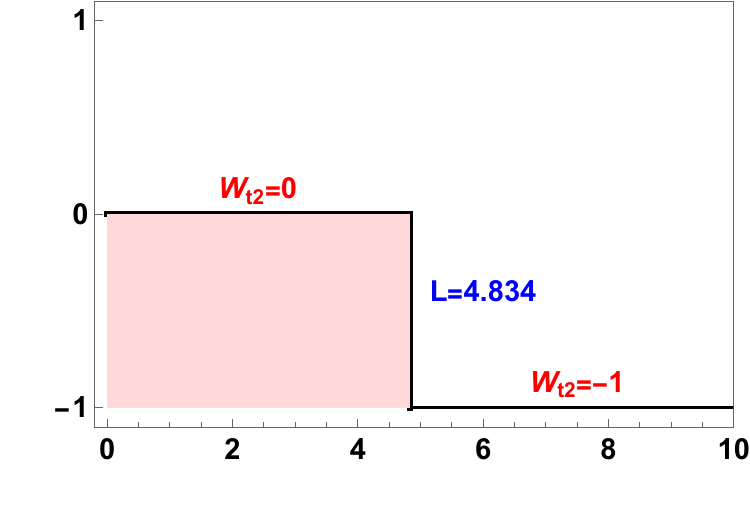}\label{TPT2b}}
    \caption{Topological phase transition of TCO within the charge regime $M<qQ<q_tQ$. The topological numbers of the TCO branches $r_{t1}$ and $r_{t2}$ change from $W_{t1}=-1$ to 0 and from $W_{t2}=0$ to -1 at $L=4.834$, respectively.} \label{TPT2}
\end{figure}

Before closing this section, we would like to give a summary on topological number $W_{RN}$ within distinct $qQ$ regimes via considering valid energy $E_{tco}$ and angular momentum $L_{tco}$. The completely results are collected into Table \ref{tab2}. From this table, we notice that the unique discrepancy between $W_{RN}$ and $W_{tot}$ shown in Tables \ref{tab2} and \ref{tab1} is that $W_{RN}=0\neq W_{tot}=-1$ in the regime of unlike strong charge. The reason for it is that if we only focus on the positive energy, the branch combining with $L_{t1}$ and $L_{t2}$ presented in Fig. \ref{O1ELa} must be discarded, which corresponds to an unstable TCO with topological number $W=-1$. On the other hand, we find that $E_{tco}$ and $L_{tco}$ can consist of three possible situations: (1) $(E_{t1},L_{t1})$; (2) $(E_{t2},L_{t2})$; (3) the combination of $(E_{t1},L_{t1})$ and $(E_{t2},L_{t2})$. According to our previous investigation, only the last situation (3) can exhibit topological phase transition, as displayed in Figs. \ref{S1ELc} and \ref{S2ELc}, which correspond to the charge regimes: $0<qQ<M$ and $M<qQ<q_tQ$, respectively. The specific details of the topological phase transitions that occur in these two charge regimes can be found in the Figs. \ref{TPT1} and \ref{TPT2}.

\begin{table}[htb]
\centering
\caption{The elements of valid energy $E_{tco}$ and angular momentum $L_{tco}$ of TCO. within the distinct $qQ$ regimes and the corresponding topological numbers $W_{RN}$ for RN black holes. Here ``PT'' stands for whether the TCO occurs phase transition.}
\vspace{3mm}
\label{tab2}
\resizebox{0.8\columnwidth}{!}{%
\begin{tabular}{cccccc}
\hline
    & \multirow{2}{*}{$qQ<-M<0$\quad} & \multirow{2}{*}{$-M<qQ<0$\quad} & \multirow{2}{*}{$0<qQ<M$\quad} & \multicolumn{2}{c}{$M<qQ<\infty$}  \\ \cline{5-6}
    &  &  &  & $M<qQ<q_tQ$ \quad  & $q_tQ<qQ<\infty$     \\ \hline
$E_{tco}$  & $E_{t1}$  & $E_{t1}$  & $E_{t1}\cup E_{t2}$  &  $E_{t1}\cup E_{t2}$  &  $E_{t2}$    \\
$L_{tco}$  & $L_{t1}$  & $E_{t1}$  & $L_{t1}\cup L_{t2}$  &  $L_{t1}\cup L_{t2}$  &  $L_{t2}$    \\ \hline
PT &No& No& Yes&Yes& No\\\hline
\multirow{2}{*}{$W_{RN}$} & \multirow{2}{*}{0} & \multirow{2}{*}{0} & \multirow{2}{*}{0} & -1 & -1 \\ \cline{5-6}
 &  &  &  & \multicolumn{2}{c}{-1} \\ \hline
\end{tabular}%
}
\end{table}

\section{Conclusions}\label{sec5}

In this paper, we mainly focused on studying the influence of the charge-to-mass ratio $q$ of massive charged test particles on the topological number  of TCOs within a static spherically symmetric spacetime with the electromagnetic potential $A_\mu=(Q/r,0,0,0)$. It is important to emphasize that in the case of neutral particles ($q=0$) within the same spacetime, the topological number is $W=0$,  implying that either no TCOs occur or TCOs appear in pairs with opposite topological charges. However, we here mainly aimed at the case $q\neq 0$, which shows a different behavior from the neutral case.

Starting with a generic Lagrangian for a massive charged test particle in this spacetime, we solved the equation of motion and obtained the effective potential $V_{\text{eff}}$, which is a quadratic function of energy. Using the decomposed energy $E_1$, we constructed a characteristic vector field $\vec{\phi}_1=(\phi^r_1, \phi^\theta_1)$ that can describe the topological properties of the TCOs. Subsequently, in order to obtain the topological number, we examined the boundary behavior of $\vec{\phi}_1$ in the region $\Sigma=[r_h,\infty)\times[0,\pi]\subset \mathbb{R}^2$. The results reveal that the vector arrows of $\vec{\phi}_1$ point upwards, downwards at $\theta=\pi$ and $0$, and to the right at $r=r_h$. These directions are independent of the charge, and are uniquely determined depicted in Fig. \ref{Bca}. However, the direction of the vector at $r\to \infty$ closely depends on the sign of $M-qQ$. When it is positive, the topological number is $W_1=0$, otherwise it is $W_1=-1$ for $E_1$. Besides that, we also investigated the boundary behaviors of the vector $\vec{\phi}_2$  constructed from $E_2$. In this case, the vector arrows of $\vec{\phi}_2$ at $r=\infty$ rely on the sign of $-M-qQ$, as displayed in Fig. \ref{BCb}. When $-M-qQ>0$, the topological number is $W_2=-1$, otherwise it is $W_2=0$ for $E_2$. Consequently, combining the analysis of vector $\vec{\phi}_1$ and $\vec{\phi}_2$, we arrived at our main conclusion: in a static spherically symmetric spacetime with a radial electric field, the presence of electric charge in massive particles can change the topology of TCOs from $W=0$ to $W=-1$ when $qQ>M$ or $qQ<-M$ for the decomposed energy $E_1$ or $E_2$, respectively.

According to the sign and value of $qQ$, we classified the parameter space into four distinct regimes: (a) unlike strong charge regime; (b) unlike weak charge regime; (c) like weak charge regime; (d) like strong charge regime. Furthermore, to illustrate our findings, we listed the total topological number $W_{tot}$ of TCOs within these $qQ$ regimes in Table \ref{tab1}, and considered RN black holes as an example. For each regime, we carefully examined the topological configuration of the unit vector $\vec{n}_1$ and $\vec{n}_2$ with respect to $E_1$ and $E_2$ in the ($r,\theta$) plane, respectively. The energy and angular momentum of TCOs are also examined. Importantly, we carried out the detailed study of the TCO radius $r_t$ and the topological number $W_{RN}$ as function of the the angular momentum $L$, via which one can obtain whether there exists topological phase transition.

In regimes (a) and (b), we identified two distinct topological configurations of the unit vector $\vec{n}_1$: $L<L_{\text{ISCO}}$ and $L>L_{\text{ISCO}}$ for $E_1$. In both cases, the topological number is $W_1=0$. When we further considered the situation of $\vec{n}_2$, the topological number $W_2=-1$ and 0 in regimes (a) and (b), respectively. That is because the configuration of $\vec{n}_2$ in regime (a) just has one possibility for any positive angular momentum. Therefore, the total topological number is $W_{tot}=W_1+W_2=-1$ and 0 for regimes (a) and (b), as presented in Table \ref{tab1}. Interestingly, we observed the existence of negative energy sectors represented by $E_{t1}$ and $E_{t2}$. By excluding them, we obtained the valid energy $E_{\text{TCO}}$ and angular momentum $L_{\text{TCO}}$ by combining both $(E_{t1},L_{t1})$ and $(E_{t2},L_{t2})$. We found that the contribution to the topological number $W_{RN}=0$ without involving negative energy in regimes (a) and (b) only comes from $E_{t1}$ and $L_{t1}$. It is worth emphasizing that the topological number $W_{RN}=0\neq W_{tot}=-1$ in the regime (a) due to the fact that an unstable TCO branch with negative energy has to be discarded.

On the other hand, in regime (c), the topological configuration of $\vec{n}_1$ and $\vec{n}_2$ built from $E_1$ and $E_2$ is identical to that in regimes (b). Therefore, the topological number remains zero, i.e., $W_1=W_2=0$. As a result, $W_{tot}=W_1+W_2=0$ is same as that in Table \ref{tab1}. However, the difference is that both $(E_{t1},E_{t2})$ and $(L_{t1},L_{t2})$ have contribution to $E_{\text{TCO}}$ and $L_{\text{TCO}}$. This observation reveals underlying topological phase transitions in $W_{t1}(L_{t1})$ and $W_{t2}(L_{t2})$, where the topological numbers change from $W_{t1}=0$ to +1 and from $W_{t2}=0$ to -1 at $L=10.709$, respectively. This nontrivial feature constitutes a significant finding in this study. In the regime (c), $W_{RN}=0$ is consistent with the result of $W_{tot}$.

In the last regime (d), the topological number is $W_{tot}=-1$. We observed a single type of topological configuration for $\vec{n}_1$ against $E_1$, which represents an unstable TCO branch with $W_1=-1$ for arbitrary $L>0$. However, regard to  $E_2$ the configuration of $\vec{n}_2$ has two kinds with $W_2=0$ according to $L_{ISCO}$. It is important to note that regime (d) needs to be further divided into two subcases: $M<qQ<q_tQ$ and $q_tQ<qQ<\infty$. In the former subcase, $E_{\text{TCO}}$ and $L_{\text{TCO}}$ are composed of $(E_{t1},L_{t1})$ and $(E_{t2},L_{t2})$. Similarly to regime (c), a topological phase transition occurs in $W_{t1}(L_{t1})$ and $W_{t2}(L_{t2})$, resulting in a change of the topological numbers from $W_{t1}=-1$ to 0 and from $W_{t2}=0$ to -1 at $L=4.834$. However, the topological number $W_{RN}=-1$ remains unchanged. In the latter subcase, $L_{t1}$ becomes an imaginary number. Thus, $E_{t1}$ and $L_{t1}$ have no contribution to $E_{\text{TCO}}$ and $L_{\text{TCO}}$. Furthermore, $W_{RN}=-1$ is in accordance with the first subcase. Our study shows that no topological phase transition occurs in this case, and the outcome of $W_{RN}=-1$ equals to $W_{tot}$ exhibited in Table \ref{tab1}. Meanwhile, all the topological number $W_{RN}$ focusing on positive energy within distinct $qQ$ regimes are organized in Table \ref{tab2}.

In conclusion, we have presented a comprehensive analysis of the topological structures of TCOs in the presence of test particles with nonvanishing charge. Our study has revealed a significant influence of the charge on the topological number. Furthermore, there are other issues for further exploration, such as extending the analysis to spinning black hole backgrounds and considering different forms of the electromagnetic potential. These directions offer promising opportunities to further understand the topology of TCOs.

\section*{Acknowledgements}
This work was supported by the National Natural Science Foundation of China (Grants No. 12075103, No. 12247101).

\end{document}